\documentclass[aps,showpacs,nofootinbib,superscriptaddress]{revtex4-1}
\usepackage{epsf}
\usepackage{bm}
\usepackage{graphicx}
\usepackage{amsfonts}
\usepackage{slashed}
\usepackage[mathscr]{euscript}
\usepackage{amsmath}
\usepackage{color}
\usepackage[normalem]{ulem}
 
\usepackage{hyperref}
\hypersetup{
    colorlinks=true,
    linkcolor=blue,
    filecolor=magenta,      
    urlcolor=cyan,
    citecolor=blue,
}
\newcommand{\stkout}[1]{\ifmmode\text{\sout{\ensuremath{#1}}}\else\sout{#1}\fi}

\def\beq{\begin{equation}}
\def\eeq{\end{equation}}
\def\be{\begin{eqnarray}}
\def\ee{\end{eqnarray}}

\def\gtap{\ \raise.3ex\hbox{$>$\kern-.75em\lower1ex\hbox{$\sim$}}\ }
\def\ltap{\ \raise.3ex\hbox{$<$\kern-.75em\lower1ex\hbox{$\sim$}}\ }

\begin{document}
\title{Polarization of tau in quasielastic (anti)neutrino scattering: the role of spectral functions}
\author{J.E. Sobczyk}\affiliation{Instituto de F\'\i sica Corpuscular (IFIC), Centro Mixto
	CSIC-Universitat de Val\`encia, Institutos de Investigaci\'on de
	Paterna, Apartado 22085, E-46071 Valencia, Spain}
\author{N. Rocco}
\affiliation{Physics Division, Argonne National Laboratory, Argonne, Illinois 60439, USA}
\affiliation{Theoretical Physics Department, Fermi National Accelerator Laboratory, P.O. Box 500, Batavia, IL 60510, USA}
\author{J.~Nieves}
\affiliation{Instituto de F\'\i sica Corpuscular (IFIC), Centro Mixto
CSIC-Universitat de Val\`encia, Institutos de Investigaci\'on de
Paterna, Apartado 22085, E-46071 Valencia, Spain}


\begin{abstract}
We present a study of the $\tau$ polarization in charged-current quasielastic (anti)neutrino-nucleus scattering.
The spectral function formalism is used to compute the differential cross section and the polarization components
for several kinematical setups, relevant for neutrino-oscillation experiments. The effects of the nuclear
corrections in these observables  are investigated by comparing the results obtained using two different realistic spectral 
functions, with those deduced from the relativistic global Fermi gas model, where only statistical correlations are accounted for. 
We show that the spectral functions, although they play an important role when predicting the differential cross sections, produce much less visible 
effects on the polarization components of the outgoing $\tau$.

\end{abstract}

\maketitle
%
%
%
%
%
%
%
%
\section{Introduction}

In the neutrino studies, the $\nu_\tau$ is experimentally the least explored one among the three neutrino flavours. Its measurement is demanding since the $\tau$ lepton, being the product of the $\nu_\tau$ charge-current (CC) interaction with matter, decays rapidly making its clear identification very challenging. There are very few $\nu_\tau$ (high energetic) events recorded. They were detected via CC interaction in OPERA~\cite{Agafonova:2014bcr} and DONuT~\cite{Kodama:2007aa} experiments.  In the near future the SHiP facility~\cite{Bonivento:2018dnp} will start operating, with the ability of measuring the cross sections of $\nu_\tau$ and $\bar\nu_\tau$ with statistics 100 times larger than the DONuT experiment.

One of the advantages of exploring $\nu_\tau$ and $\bar\nu_\tau$ CC interactions is the fact that for a wide range of energies, the outgoing $\tau$ is not fully polarized. Each of the $\tau^-$ and $\tau^+$ polarization components (longitudinal and transverse ones) is sensitive to different combinations of the nuclear structure functions, making them interesting observables to further explore the differences between various nuclear models. They convey richer information, which is complementary to the knowledge obtained by means of the cross section predictions. In the limit of high energies $E_\nu \gg m_\tau$, though, the outgoing $\tau$ leptons are produced in totally polarized states. Thus, the interesting energy region to be explored is limited to the values of $E_\nu \lesssim 10$ GeV. This relatively moderate energy range can be studied by oscillation experiments, although the measurement is demanding because of the low statistics. From this perspective, the information about the final $\tau$ lepton polarization could be helpful because its spin direction affects the angular distribution of the decay products.

We will focus on the quasi-elastic (QE) region in which the single nucleon knock-out is the dominant reaction mechanism. Previous works on $\nu_\tau$ and $\bar\nu_\tau$ scattering have considered the nucleus as an ensemble of free nucleons~\cite{Hagiwara:2003di}, or used the Random Phase Approximation (RPA), and 
an effective nucleon mass to describe the initial nuclear state~\cite{Graczyk:2004uy}. The use of an effective mass for the nucleon is a simplified method to account for the effects due to the change of its dispersion relation inside of a nuclear medium. A proper description, however, is achieved by dressing the nucleon propagators and constructing realistic particle and hole spectral functions (SFs), which incorporate dynamical effects  that depend on both the energy and momentum of the 
nucleons~\cite{Nieves:2017lij}.  There was an attempt to include the SF formalism in the study  of the polarization of the outgoing lepton produced in CC (anti-)neutrino-nucleus reactions~\cite{Valverde:2006yi}. However, because of the non-relativistic nature of the nuclear calculations carried out in that study, the predictions were restricted to a very narrow region of the available phase space. In this work we perform an analysis which does not suffer from the above-mentioned problem, and we use realistic hole SFs 
to obtain the $\tau-$polarization vector in the whole available phase-space for neutrino energies below 10 GeV.  Moreover, to gauge the model-dependence of the predictions, we perform the calculation using spectral functions obtained within two theoretically different frameworks: a semi-phenomenological one 
based on the findings of Ref.~\cite{FernandezdeCordoba:1991wf}, and a second one~\cite{Benhar:1989aw,Benhar:1994hw} obtained within the 
Correlated Basis Function (CBF) theory~\cite{Fantoni:1987hfr}. Both sets of SFs provide a realistic description of the dynamics including nucleon-nucleon correlations in the initial target and in the 
nuclear remnant. Moreover, combined with a factorization scheme, they have been successfully used for modeling inclusive electro-- and (anti)neutrino--nuclear 
QE responses~\cite{Gil:1997bm, Nieves:2004wx, Nieves:2005rq, Nieves:2017lij, Sobczyk:2017mts, Benhar:2005dj, Benhar:2006nr, Benhar:2009wi, Benhar:2010nx, Benhar:2013dq, Rocco:2015cil, Vagnoni:2017hll}.

The lepton polarization results presented here, and the comparison for these observables 
of the SF models of Ref.~\cite{FernandezdeCordoba:1991wf} and Ref.~\cite{Benhar:1989aw, Benhar:1994hw}, are a natural continuation 
of  the works of Refs.~\cite{Sobczyk:2017vdy} and \cite{Sobczyk:2019uej}, where the electroweak scaling properties and  the strange hyperon production in nuclei, respectively, were studied.  

Finally, we would also like to mention that the inclusion of RPA correlations  do not change appreciably 
the gross features of the polarization of the $\tau$'s. The reason is
that the polarization components are obtained as a ratio between linear combinations of nuclear structure functions and the RPA changes similarly numerator and denominator. In addition, one should bear in mind that RPA corrections take into account the absorption of the gauge boson, mediator of the interaction, by the nucleus as a whole instead of by an individual nucleon, and their importance decreases as the gauge boson wave-length becomes much shorter 
than the nuclear size. Thus, RPA effects on the polarization observables become little relevant, even for the total or partially integrated cross sections~\cite{Gran:2013kda, Nieves:2017lij}, 
 for a great part of the phase-space accessible in the CC reaction~\cite{Graczyk:2004uy, Valverde:2006yi}. 

This work  is organized as follows. In Sec. \ref{sec:formalism} we introduce the formalism and the basic concepts. In Subsec.~\ref{sec:hadron_tensor} we focus our attention on the hadron tensor, using the spectral function formalism to rewrite it. In Sec.~\ref{sec:results}, in the first place, we define the phase space to be explored in our analysis, and then perform a comparison of the results obtained using two realistic SF models  and the more approximate Relativistic Global Fermi gas (RGFG) approach. Finally, in Sec.~\ref{sec:conclusions} we draw our conclusions, while some details on the definition of the lepton polarized CC cross section and  some kinematical relations for weak charged lepton production off nucleons are given in the Appendices~\ref{lept:hadr:terms} and \ref{sec:app}, respectively.

%
%
%
%
%
%
%
%
\section{Formalism}\label{sec:formalism}
We will investigate CC (anti-)neutrino--nucleus reactions
\begin{equation}
\nu_\ell/\bar\nu_\ell(k) + A \to \ell^{\mp}(k') + X\, ,\quad \ell=e,\mu,\tau
\end{equation}
where $k^\mu$, $k'^\mu$ are the  incoming- and outgoing-lepton four momenta, respectively.

\subsection{Charged lepton polarization}

The (anti)neutrino inclusive-differential cross section  for a $(s;h)-$polarized  outgoing lepton can be 
written using Eqs.~(\ref{eq:1}) and (\ref{eq:lept_pol}) of the appendix as
\begin{eqnarray}
 \left.\frac{d^2\sigma^{(\nu_\ell, \bar\nu_\ell)}}{d\Omega (\hat{k}') d E_{k'}}\right|_{s;h} &=& \frac12\Sigma_0^{(\nu_\ell, \bar\nu_\ell)}\left( 1 + h\, s_\alpha {\cal P}^\alpha_{(\ell^-,\, \ell^+)}\right)  \label{eq:sec-pol}\\ 
 \Sigma_0^{(\nu_\ell, \bar\nu_\ell)} &=& \frac{|\vec{k}'|\, G_F^2 M_i}{2\pi^2} F^{(\nu_\ell, \bar\nu_\ell)} \label{eq:sec-unpol}
\end{eqnarray}
where $\Sigma_0$ is the LAB double differential cross section corresponding to unpolarized leptons and ${\cal P}^\mu$ is the polarization vector. The term $F$ depends on the lepton and hadron kinematics
as discussed more in detail in Appendix~\ref{lept:hadr:terms}. 
In the hadronic tensor, only five out of the six structure functions, $W_i$, contributes to $F$~\cite{Nieves:2004wx}:
\begin{eqnarray}
\frac{W^{\mu\nu}_{(\nu_\ell, \bar\nu_\ell)}}{2 M_i} &=& -g_{\mu\nu} W_1^{(\nu_\ell, \bar\nu_\ell)} + \frac{P^\mu P^\nu}{M_i^2}W_2^{(\nu_\ell, \bar\nu_\ell)}+ i\frac{ \epsilon^{\mu\nu\alpha\beta}P_{\alpha} q_\beta}{2 M_i^2}W_3^{(\nu_\ell, \bar\nu_\ell)} +  \frac{q^\mu q^\nu}{M_i^2}W_4^{(\nu_\ell, \bar\nu_\ell)}+ \frac{P^\mu q^\nu+q^\mu P^\nu}{2 M_i^2}W_5^{(\nu_\ell, \bar\nu_\ell)}\nonumber \\
&&+i \frac{P^\mu q^\nu-q^\mu P^\nu}{2 M_i^2}W_6^{(\nu_\ell, \bar\nu_\ell)}\, .
\label{eq:hadron_general}
\end{eqnarray}
The structure functions $W_i$, are real Lorentz scalars that depend on $q^2$. They encode the nuclear response to the electroweak probe, which is determined by a variety of mechanisms, {\em e.g.} QE scattering, two-nucleon knockout, pion production, and deep inelastic scattering. 
The term proportional to $W_6$ does not contribute to the double differential cross section, and thus it does not appear in the full expression for $F$ that can be found in Eq.~(2) of Ref.~\cite{Valverde:2006yi}\footnote{Actually, the term $F$ can be also read off from Eq.~(10) of Ref.~\cite{Nieves:2004wx}. We should mention that there is a typo in this latter equation that affects  the $W_4$ term, where $\sin^2\theta$ should be $\sin^2\theta/2$, with  $\theta$  the angle between $\vec{k}$ and $\vec{k}'$ in the LAB system.}. 
When contracting the leptonic and hadronic tensor, $W_3$ contributes to $F$ with an opposite signs for the antineutrino and neutrino scattering. 
Moreover, the value of the nuclear structure functions will also be different for the two reactions. In the former case, the scattering takes place on a bound proton, {\textit e.g.} the $W_i$ 
will depend on the proton mass and removal energy, in the latter the struck nucleon is a neutron. 
The differences between the antineutrino and neutrino results are expected to be particularly significant in asymmetric nuclei, such as  $^{40}$Ar. 

The polarization vector ${\cal P}^\alpha$ of the outgoing $\ell^{\mp}$ lepton is determined by the nuclear response,
\begin{eqnarray}
 {\cal P}_\alpha^{(\ell^-,\, \ell^+)} &=& \mp m_\ell \frac{\left( k_\mu g_{\nu\alpha}+ k_\nu g_{\mu\alpha}-k_\alpha g_{\mu\nu} \pm i\epsilon_{\mu\nu\alpha\beta}k^\beta\right)W^{\mu\nu}_{(\nu_\ell, \bar\nu_\ell)}}{[L_{\mu\nu}W^{\mu\nu}]_{(\nu_\ell, \bar\nu_\ell)}} \,,  \label{eq:vecpol}
\end{eqnarray}
which  can be decomposed as follows 
\begin{equation}
 {\cal P}^\alpha_{(\ell^-,\, \ell^+)} =  - \left(P_L n^\alpha_l + P_T  n^\alpha_t +  P_{TT}  n^\alpha_{tt}\right)\big|_ {(\ell^-,\, \ell^+)} \label{eq:def-pol-vector}
\end{equation}
where the three four-vectors $n_l$, $n_t$ and $n_{tt}$ are given by
\begin{equation}
 n^\alpha_l = \left(\frac{|\vec{k}'|}{m_\ell}, \frac{E_{k'}\vec{k}'}{m_\ell|\vec{k}'|} \right), \qquad n^\alpha_t = \left(0, \frac{(\vec{k}\times\vec{k}')\times \vec{k}'}{|(\vec{k}\times\vec{k}')\times \vec{k}'|}\right), \qquad n^\alpha_{tt} = \left(0, \frac{\vec{k}\times\vec{k}'}{|\vec{k}\times\vec{k}'|}\right)\, \label{eq:enes}
\end{equation}
We have ignored the projection of 
${\cal P}^\alpha $  onto the direction of the four vector $k'^\alpha$, because it is irrelevant for the $(s;h)-$polarized differential cross section since
$s\cdot k'=0$. In addition, ${\cal P}\cdot n_{tt}=0$ and therefore $P_{TT}=0$,  which means that the polarization
three-vector lies in the lepton-scattering plane (see Fig.1 of Ref.~\cite{Valverde:2006yi}). Note that under parity, ${\cal P}^\alpha $ transforms as
\begin{equation}
 {\cal P}^\alpha \to {\cal P}_\alpha 
\end{equation}
which automatically requires $P_{TT}=0$, since $n^\alpha_{tt}$ stays invariant under a parity transformation. (Time reversal invariance can be also used to show that $P_{TT}=0$). In addition, it is obvious that ${\cal P}^2$, called the degree of polarization~\cite{Hagiwara:2003di}, 
\begin{equation}
-{\cal P}^2_{(\ell^-,\, \ell^+)} = (P_L^2 + P_T^2)\big|_{(\ell^-,\, \ell^+)} 
\end{equation}
is a Lorentz scalar, as $P_{L,T}$ also are, since they can be computed taking scalar products. i.e.,  $P_{L,T}=-({\cal P}\cdot n_{l,t})$. From the polarized double differential cross section of Eq.~(\ref{eq:sec-pol}), we obtain these longitudinal and perpendicular components 
of the outgoing lepton polarization vector as follows
\begin{equation}
 P_{L,T}^{(\ell^-,\, \ell^+)} = \frac{\left.\frac{d^2\sigma^{(\nu_\ell, \bar\nu_\ell)}}{d\Omega (\hat{k}') d E_{k'}}\right|_{n_{l,t}}^{h=+}-\left.\frac{d^2\sigma^{(\nu_\ell, \bar\nu_\ell)}}{d\Omega (\hat{k}') d E_{k'}}\right|_{n_{l,t}}^{h=-}}{\left.\frac{d^2\sigma^{(\nu_\ell, \bar\nu_\ell)}}{d\Omega (\hat{k}') d E_{k'}}\right|_{n_{l,t}}^{h=+}+\left.\frac{d^2\sigma^{(\nu_\ell, \bar\nu_\ell)}}{d\Omega (\hat{k}') d E_{k'}}\right|_{n_{l,t}}^{h=-}}  = \frac{1}{\Sigma_0^{(\nu_\ell, \bar\nu_\ell)}}\left\{\left.\frac{d^2\sigma^{(\nu_\ell, \bar\nu_\ell)}}{d\Omega (\hat{k}') d E_{k'}}\right|_{n_{l,t}}^{h=+}-\left.\frac{d^2\sigma^{(\nu_\ell, \bar\nu_\ell)}}{d\Omega (\hat{k}') d E_{k'}}\right|_{n_{l,t}}^{h=-}\right\} \label{eq:assy}
\end{equation}
By construction, it follows that $|P_{L,T}| \le 1$. Furthermore, $(P_L^2 + P_T^2)\le 1$, because $|{\cal P}^2\,|\le 1$. This can be easily deduced in the outgoing lepton rest frame, considering that in this system, $|\vec{\cal P}\cdot\hat n|\le 1$ for any unit vector $\hat n$, since for both polarizations $h=\pm$, the double differential cross section $\left.\frac{d^2\sigma^{(\nu_\ell, \bar\nu_\ell)}}{d\Omega (\hat{k}') d E_{k'}}\right|_{\hat n;h}\ge 0$.

The $P_{L,T}$ components depend on the lepton kinematics and on the structure functions, $W_i$, introduced in Eq.~(\ref{eq:hadron_general}). Explicit expressions for these observables in the LAB system are given in Eqs.~(5) and (6) of Ref.~\cite{Valverde:2006yi}, which were obtained from the findings of \cite{Hagiwara:2003di}. Besides masses, they depend on the scalars $(k\cdot P)$, $(k'\cdot P)$ and $q^2$, which define the neutrino and outgoing lepton energies and the angle, $\theta$, 
between $\vec{k}$ and $\vec{k}'$ in the LAB system.

It can be seen that for $W_3=0$ , $\ell^+$  and $\ell^-$ have opposite polarizations, up to some effects due to the asymmetry of the role played by protons and neutrons in the  nuclear system.

The operators $(1\pm \gamma_5 \slashed{n_l})/2$, with $n_l^\alpha$ obtained from $s^\alpha$ in Eq.~(\ref{eq:smu}) using $\vec{k}'/|\vec{k}'|$ as unit vector, are  helicity projectors~\cite{Mandl:1985bg}, and thus, the asymmetry proposed in Eq.~(\ref{eq:assy}) for the case of  $P_L$ turns out to be the  outgoing lepton helicity asymmetry. Moreover, since at high energies helicity and chirality  coincide, and the latter is conserved in CC reactions, we conclude
\begin{equation}
\lim_{(m_\ell/|\vec{k}'|)\to 0} P_L^{\ell^-} = -1\, , \qquad \lim_{(m_\ell/|\vec{k}'|)\to 0} P_L^{\ell^+} = 1\, ,
\end{equation}
which follows from the negative (positive) chirality of the neutrino (antineutrino) that is inherited by the outgoing $\ell^- (\ell^+)$ produced in the CC transition. In addition, in the $(m_\ell/|\vec{k}'|)\to 0$ limit, the transverse polarization ${\cal P}_T$ vanishes, for both neutrino and antineutrino processes. Indeed, it is proportional to the outgoing lepton mass and to $\sin\theta$ \cite{Valverde:2006yi, Hagiwara:2003di}. (Note that for $P_L$, $m_\ell$ in the definition of ${\cal P}^\alpha$ in Eq.~(\ref{eq:vecpol}) cancels out with the $1/m_\ell$ common factor that contains $n^\alpha_l$ in Eq.~(\ref{eq:enes})). In this ultra-relativistic energy regime the whole non-trivial behaviour of the polarization components, coming from the hadron tensor,  cancels out in the ratio taken in Eq.~\eqref{eq:vecpol}.

In the case of electron and muon CC production the cross section depends mostly on $W_1$, $W_2$ and $W_3$, while the contribution of the other structure functions are suppressed by the small lepton mass. Therefore, $P_T$ takes small values close to zero, while  $P_L$ is expected to differ little from the asymptotic $\mp 1$ values for neutrino or antineutrino reactions, respectively, in most of the available phase  space.

%
%
%
%
%
%

\subsection{QE Hadron tensor}
\label{sec:hadron_tensor}
\subsubsection{Vacuum}

When considering the interaction of a (anti-)neutrino with a single free nucleon of momentum $p$ and mass $M$,
\begin{equation}
 \nu_{\ell} + n \to p+\ell^-\, , \qquad  \bar\nu_{\ell} + p \to n+\ell^+
\end{equation}

the hadron dynamics is determined by the  nucleon tensor $A^{\mu\sigma}$ (we use the same conventions as in Ref.~\cite{Sobczyk:2019uej}):
\begin{align}
A^{\mu\sigma}(p,q)&= \overline\sum \langle p+q |  j^\mu_{cc\pm}(0) | p \rangle\langle p+q |  j^\sigma_{cc\pm}(0) | p \rangle ^*= \overline\sum \big[ \bar{u}(p+q) (V^\mu -A^\mu) u(p) \big]  \big[ \bar{u}(p+q)(V^\sigma-A^\sigma) u(p) \big]^* \nonumber \\
&=  \frac12 {\rm Tr}\left[ \frac{(\slashed{p}+\slashed{q}+ M)}{2M} (V^\mu -A^\mu)\frac{(\slashed{p}+M)}{2M}\gamma^0(V^\sigma-A^\sigma)^\dagger
\gamma^0\right] \label{eq:amunu}
\end{align}
with $M$ and $u, \bar u$,  the nucleon mass and  the dimensionless Dirac nucleon spinors, respectively, and we have summed and averaged over initial and final nucleon spins. In addition, 
\begin{equation}
V^\mu = 2\cos\theta_C \times  \left ( F_1^V(q^2)\gamma^\mu + {\rm
i}\mu_V \frac{F_2^V(q^2)}{2M}\sigma^{\mu\nu}q_\nu\right), \qquad
A^\mu = \cos\theta_C G_A(q^2) \times \left (  \gamma^\mu\gamma_5 + 
\frac{2M}{m_\pi^2-q^2}q^\mu\gamma_5 \right) 
\end{equation}
with $m_\pi=139$ MeV, the  pion mass\footnote{Note that the CC nucleon tensor, $\widetilde{A}^{\mu\nu}$, defined in Eq.~(27) of Ref.~\cite{Nieves:2004wx} is related to that given here in Eq.~(\ref{eq:amunu}) by  $\widetilde{A}^{\mu\sigma}(p,q) = 8M^2 A^{\sigma\mu}(p,q)/\cos^2\theta_C$. In addition, $\widetilde{A}^{\mu\nu}$ is explicitly given in the Appendix A of that reference for on-shell nucleons. There is a typo there that affects the overall sign of the $xy$ component (Eq.~(A8)). }. Partial conservation of the  axial current  and
invariance under G-parity have been assumed to relate the
pseudoscalar form factor to the axial one and to discard a term of the
form $(p^\mu+p^{\prime \mu})\gamma_5$ in the axial sector,
respectively.  While the pseudoscalar form factor can be safely neglected when considering $\nu_e$, $\nu_\mu$-induced processes, its contribution, proportional to $m^2_\ell$, becomes more relevant for $\tau$ production.  Invariance under time reversal guarantees that all form
factors are real, and  besides, due to isospin symmetry, the vector
form factors are related to the electromagnetic ones.  

We use in this work the parametrizations  of the form-factors given in \cite{Nieves:2004wx}, which were also employed in the Ref.~\cite{Valverde:2006yi} (Galster et al.~\cite{Galster:1971kv} in the vector sector and a dipole form for the axial one). 

\subsubsection{Nuclear medium}

For sufficiently large values of the momentum transfer ($|\vec{q}\,|\gtrsim 500$ MeV) the lepton-nucleus scattering can be safely treated within the Impulse Approximation (IA). In this framework the hadron tensor of Eq.~\eqref{eq:2} is obtained as a convolution of the spin averaged squared
amplitude of the hadron matrix element of Eq.~\eqref{eq:amunu} and the hole SF:
\begin{align}
W^{\mu\nu}(q)=& \Theta(q^0) \int \frac{d^3p}{(2\pi)^3} \int_{\mu-q^0}^\mu dE S_h(E, \vec{p}\,)\frac{M^2}{E_p\,E_{p+q}} A^{\mu\nu}(p,q) \delta(q^0+M+E-E_{p+q})\, .
\label{had:tens}
\end{align}
where $\mu$ is the chemical potential and  the initial nucleon is considered off-shell. 
The factors  $M/E_p$ and $M/E_{p+q}$, with $E_p=\sqrt{M^2+\vec{p}\,^2}$ are included to 
account for the implicit covariant  normalization of the Dirac spinors of the initial and final nucleons in the matrix elements 
of the relativistic current in Eq.~\eqref{eq:amunu}.
The hole SF, $S_h(E,\vec{p})$, encompasses information on the internal nuclear structure, giving
the probability distribution of removing a nucleon with momentum $\vec{p}$ from the target nucleus, 
leaving the residual $(A-1)-$nucleon system with an excitation energy $-E \ge 0$. For simplicity, we consider a 
symmetric nucleus with equal proton and nucleon density distributions $\rho_n=\rho_p=\rho/2$.

In this work we use three different nuclear models to obtain the hole SF. 

\begin{itemize}

\item In the RGFG model only statistical (Pauli) correlations among the nucleons are accounted for, yielding
to a very simple expression 
\begin{align}
S_h^{\text{RGFG}}(E,\vec{p}\,)&=\frac{2A}{\bar\rho}\delta[E+M-(E_p-\epsilon)]\theta(|\vec{p}\,|-\bar{p}_F)\ , 
\end{align}
where $\bar\rho=2\bar{p}_F^3/(3\pi^2)$ is the averaged total nucleon (both proton and neutron) density, $A$ is the total number of nucleons
and $\sqrt{M^2+{\bar p}_F^2}$ is the chemical potential in this model.  We approximate the shift-energy $\epsilon>0$ by the difference between the
experimental masses of the ground-states of the initial and final nuclear systems. To account for it, we  use a shifted value of $q^0\to (q^0-\epsilon)$ 
(see Ref.~\cite{Nieves:2004wx} for more details, where for a isospin symmetric nuclear matter, $\epsilon$ is identified with the $Q-$value of the reaction). The effective Fermi momenta $\bar{p}_F$ are determined from the analysis of electron-scattering data of Ref.~\cite{Maieron:2001it} for different nuclear species. In this work, we will show results for $^{16}$O, for which we take $\bar{p}_F=225$ MeV and $\epsilon\sim 15 (11)$ MeV for $\nu_\ell$ ($\bar\nu_\ell$) induced reactions. 

It has long been known that the RGFG method is oversimplified and that a realistic description of nuclear dynamics, including correlations among the nucleons, 
is needed to provide an accurate description of  electroweak scattering off nuclei. 
We consider two nuclear SFs, derived within the framework of nuclear many-body theory using the CBF approach~\cite{Benhar:1989aw,Benhar:1994hw} and the semi-phenomenological 
one of Ref.~\cite{FernandezdeCordoba:1991wf}, that starts from the experimental elastic nucleon-nucleon cross section, with certain medium corrections. Both approaches  use the Local Density Approximation (LDA) to obtain finite nuclei results from those calculated in nuclear matter, and despite being based on different models of nuclear dynamics, provide compatible nucleon-density scaling functions~\cite{Sobczyk:2017vdy}.

\item In the formalism of Ref.~\cite{FernandezdeCordoba:1991wf}, used for neutrino QE scattering in Refs.~\cite{Nieves:2004wx, Nieves:2017lij}, one  firstly  performs a calculation for the nuclear medium at a certain density, and then integrate it over the density profile $\rho(r)$ of a given nucleus. Hence,  it fully relies on the LDA and, in what follows,  we will denote by LDA the physical properties and the results obtained within this scheme.

The energy-momentum distribution of nucleons in the symmetric nuclear matter of density $\rho$ can be described by means of hole and particle SFs  determined by the nucleon self-energy in the medium (see 
Refs.~\cite{Nieves:2004wx, Nieves:2017lij} and in particular the Section III of \cite{Sobczyk:2017vdy})
\beq
S_{p,h}^{\text{LDA}}(E, \vec{p}\,) = \mp \frac{1}{\pi} \frac{\text{Im}\Sigma(\widehat{E}, \vec{p}\,)}{\big(E- \vec{p}\,^{2}/2M - \text{Re}\Sigma(\widehat{E}, \vec{p}\,)-C\rho \big)^2 + [\text{Im}\Sigma(\widehat{E}, \vec{p}\,)]^2}, \quad  \widehat{{\cal C}}= \text{Re}\Sigma(p_F^2/2M,p_F) + C\rho
\eeq
with the chemical potential given by $\mu = p_F^2/2M + \widehat{{\cal C}}$, $\widehat{E}=E-\widehat{{\cal C}}$, $E\le \mu $ or $E > \mu $ for $S_h$ or $S_p$, respectively, and $p_F=(3\pi^2\rho/2)^\frac13$. Since $\text{Im}\Sigma(E, \vec{p}\,) \ge 0 $ for $E\le\mu$, and $\text{Im}\Sigma(E, \vec{p}\,) \le 0 $ for $E > \mu$, the chemical potential can be defined as the point 
in which $\text{Im}\Sigma(E,  \vec{p}\,)$ changes sign.  In addition, the dependence of the nucleon self-energy, $\Sigma(\widehat{E}, \vec{p}\,)$, on $\rho$ 
is implicit. 

Both real and imaginary parts of $\Sigma$ are obtained from the semi-phenomenological model 
derived in \cite{FernandezdeCordoba:1991wf}, which,  starting from the experimental elastic $NN$ scattering cross section, incorporates, consistently  with the low density theorems, some medium polarization (RPA) corrections. The approach is non-relativistic and it is derived for isospin symmetric nuclear matter. The resulting nucleon self-energies stay in a good agreement with microscopic calculations \cite{Fantoni:1983zz, Fantoni:1984zz, Ramos:1989hqs, Muther:1995bk, Mahaux:1985zz}, and provide  effective masses, nucleon momentum distributions, etc., which are also in good agreement with
sophisticated many-body results~\cite{Ramos:1989hqs, Mahaux:1985zz}. 

The real part of the self-energy calculated in \cite{FernandezdeCordoba:1991wf} 
should not be treated as an absolute value, since  momentum independent terms are not considered there, and it should be understood 
as an energy difference from $ \text{Re}\Sigma(p_F^2/2M, p_F)$. This, in principle,  does not prevent the approach to be used to compute particle-hole (QE) response functions, where only differences between two nucleon self-energies appear, and  the  constant terms  of the hole and particle self-energies cancel~\cite{Nieves:2017lij}.  However, the use of non-relativistic kinematics is sufficiently  accurate for the hole, but its applicability to the ejected nucleon limits the range of energy and momentum transferred to the nucleus. For high energetic ejected nucleons and inclusive cross sections,  the IA is a good approximation, as mentioned in the beginning of this section\footnote{For instance, see the discussion in Ref.~\cite{Nieves:2017lij} of the 
results for the CCQE cross section $\sigma(\nu_\mu+ ^{12}{\rm C} \to \mu^- + X)$ obtained in \cite{Vagnoni:2017hll} within the IA.}.  Such approximation considers a fully dressed nucleon-hole, but uses a free particle SF, i.e., it employs a plane wave for the outgoing nucleon, satisfying a free relativistic 
energy-momentum dispersion relation. To obtain results using a dressed hole and an undressed particle, an absolute value for the real part of the nucleon-hole self-energy is needed. This is achieved by including the  phenomenological constant term $C\rho$ in the nucleon self-energy, with $C=0.8$ fm$^2$ for carbon, 
fixed to  a binding energy per nucleon $|\epsilon_A| = 7.8$ MeV. Some more details can be found in  Refs.~\cite{Sobczyk:2017vdy, Marco:1995vb}. We will use this value for the $C-$parameter here also for oxygen.

In this context, assuming a free outgoing on-shell nucleon, the hadron tensor for finite nuclei is  obtained as
\be \label{eq:LDA}
W^{\mu\nu}_{\text{LDA}}(q) =2\Theta(q^0)\int d^3r   \int \frac{d^3p}{(2\pi)^3}\int_{\mu-q^0}^{\mu} dE\, S_h^{\text{LDA}}(E, \vec{p}\,)\frac{M^2}{E_p\,E_{p+q}} \delta(q^0+E+M-E_{p+q})  A^{\mu\nu}(p,q)\bigg|_{p^0=E+M} 
\ee
where we have used $S_h^{\text{LDA}}(E, \vec{p}\,)$ as a function of the nuclear
density at each point of the nucleus and integrate over the whole
nuclear volume. Hence, we assume the LDA, which is an
excellent approximation for volume processes~\cite{Carrasco:1989vq},  like the one studied here. Let us notice that by setting $p^0=E+M$, we calculate $A^{\mu\nu}(p,q)$ for an off-shell nucleon, {\em i.e.} we take  the energy and momentum distributions from the hole spectral function $S^{\text{LDA}}_h(E, \vec{p}\,)$, changing the dispersion relation of the initial nucleon. However, there exists a little inconsistency here, since the sums over nucleons' spins in Eq.~\eqref{eq:amunu} were carried out assuming free dispersion relations for the nucleons. As we will see below, this procedure is  more accurate than setting $p^0 = E_p$ in Eq.~(\ref{eq:LDA}), as was done in previous works \cite{Nieves:2004wx, Valverde:2006yi, Nieves:2017lij}. Nevertheless, the differences are relatively 
small and visible only for forward scattering.

\item The CBF spectral function for finite nuclei is given by the sum of two different terms~\cite{Benhar:1994hw}
\begin{align}
S_h^{\text{CBF}}(E,\vec{p}\,)&=S^{\rm MF}_h (-E, \vec{p}\,)+S^{\rm corr}_h (-E, \vec{p}\,)\, .
\end{align}
The first one is derived from a modified mean field (MF) calculation, where experimental information, obtained from $(e,e'p)$ scattering data, are used
to account for residual effects of nucleon interactions neglected in a MF picture~\cite{Mougey:1976sc,Turck:1981,Dutta:2003yt}. The second term determines the behavior of the hole SF in the
high momentum and removal energy region. It has been obtained by folding CBF calculations of the SF in uniform and isospin symmetric nuclear matter with the nuclear density distribution profile ~\cite{Benhar:1989aw, Benhar:1994hw}. We remind here that $-E\ge 0$ is the excitation energy of the residual $(A-1)-$nucleon system.
Within the IA the scattering off a bound nucleon can be accounted for using different approximations. The CBF results which we  present in this work  have been obtained replacing in the one-body current operator $j^\mu_{cc\pm}$ of Eq.~\eqref{eq:amunu}, the four momentum $q^\mu=(q^0,\vec{q}\,)$ by $ \tilde{q}^{\, \mu}=(\tilde{q}^{\,0},\vec{q}\,)$, such that  $\tilde{q}^{\,0}= q^0 -(E_p-M-E)$, in analogy with the prescription adopted in the RGFG case.  The quadri-spinors entering the evaluation of the hadron tensor are those of free nucleon states. 
Note that this approximation leads to a violation of current conservation for the electromagnetic case. Different procedures aimed at restoring the gauge invariance have been discussed in Refs.~\cite{DeForest:1983ahx,Benhar:2006wy}. In particular, the authors of Ref.~\cite{Benhar:2006wy} argue that the violation of gauge invariance in the IA scheme is expected to become less and less important in inclusive electron scattering at large momentum transfers.\\

\end{itemize}

%
%
%
%
%
%
%
%
\section{Results}\label{sec:results}
The analysis carried out in Ref.~\cite{Hagiwara:2003di} clearly shows that for LAB energies of the $\nu_\tau/\bar{\nu}_\tau$ beam comprised between 3.5 and 10 GeV, the QE cross section is sizable. This study was done for a scattering on a single nucleon and neglecting multi-nucleon emission; their contribution, albeit non vanishing,  would be smaller than the former one.
 The breakdown of the total neutrino cross section into the QE, pion production, and deep inelastic scattering contributions  is shown in Fig.~5 of Ref.~\cite{Hagiwara:2003di}. The QE mechanism is found to be dominant up to $E_{\nu}\sim 6$ GeV and the same observation holds true for $\bar{\nu}_\tau$ reactions. 
 
\subsection{QE mechanism phase space} 
 The large mass of the $\tau$ lepton ($m_\tau \approx 1776.8$ MeV) greatly limits the  phase space available for the single-nucleon knockout processes, being prohibited the large LAB dispersion angles, as discussed in Appendix~\ref{sec:app}. In Fig.~\ref{fig:phase-space}, we analyze the phase space for different values of the incoming neutrino energy, $E_{\nu}$, and  of the lepton scattering  angle in the LAB system ($\theta)$. We show 
\begin{equation}
|\vec{q}\,|_{[E_\nu, q^0, \theta]} = \left[E_\nu^2+ (E_\nu-q^0)^2-m_\tau^2-2E_\nu\sqrt{(E_\nu-q^0)^2-m_\tau^2}\cos\theta\right]^\frac12 \label{eq:defqtau}
\end{equation}
as a function of the energy transfer $q^0$, together with the QE-peak curve $| \vec{q}\,|_{\rm QE}= \sqrt{2Mq^0+(q^{0})^2}$ (black solid line).  The shaded areas in Fig.~\ref{fig:phase-space}  have been obtained by varying the outgoing lepton scattering angle from $0^\circ$ to $16^\circ, 24^\circ, 28^\circ$ and $30^\circ$ for $E_{\nu}$= 4, 6, 8 and 10 GeV, respectively. All the angles chosen to evaluate the upper bands are greater than  $\theta_{\rm max} (E_\nu)$, introduced in the Appendix~\ref{sec:app}. It corresponds to the maximum allowed LAB $\tau-$scattering angle for the weak production off a free nucleon, and  its dependence on the incoming neutrino energy is given in Eq.~\eqref{eq:thetamax}. This limiting angle takes the approximate  values of $11.4^\circ$, $20.3^\circ$, $23.6^\circ$ and $25.5^\circ$ for $E_\nu=$4 , 6, 8  and 10 GeV, respectively (see also the right panel of Fig.~\ref{fig:thetaLAB} for further details). The range of $q^0$ values, $[q^0_{\rm min}, q^0_{\rm max}]$, for which there exist  solutions 
($\theta$) of the QE condition $| \vec{q}\,|_{\rm QE}=|\vec{q}\,|_{[E_\nu, q^0, \theta]}$ grows rapidly with $E_\nu$. Actually at threshold, $E_\nu= m_\tau+m_\tau^2/2M\sim 3.46$ GeV,  $q^0_{\rm min}=q^0_{\rm max}\sim 0.58$ GeV, and when the neutrino energy gets bigger,  $q^0_{\rm min}$ and $q^0_{\rm max}$ quickly decreases and increases, tending to zero and to $E_\nu-(m_\tau^2+M^2)/2M$, respectively. The $q^0-$range is shown in the left panel of Fig.~\ref{fig:thetaLAB} of the Appendix~\ref{sec:app}, up to $E_\nu = 10$ GeV. 

Additionally in each panel of Fig.~\ref{fig:phase-space}, the yellow curve shows $|\vec{q}\,|_{[E_\nu, q^0, \theta]}$ for an intermediate angle among 
those accounted for in the band. We see that this curve, as well as the $(\theta=0^\circ)-$one, intercepts twice 
the $| \vec{q}\,|_{\rm QE}-$line. This is because for any LAB $\tau-$scattering angle $\theta \le\theta_{\rm max} (E_\nu)$, there exist two different values of the LAB $\tau-$energy that satisfies the QE condition $| \vec{q}\,|_{\rm QE}=|\vec{q}\,|_{[E_\nu, q^0, \theta]}$. Thus, we might expect the existence of two QE peaks in the nuclear differential cross section, located at different values of $q^0$ for fixed $E_\nu$ and $\theta$ LAB variables. This never occurs for charged muon or electron productions, except for a extremely narrow range of neutrino energies. A detailed discussion can be found in 
Appendix~\ref{sec:app}, and in particular this non-biunivocal correspondence between $\tau-$lepton LAB scattering angle and energy is
shown in the right panel of Fig~\ref{fig:thetaLAB}.  

One should  bear in mind that the nuclear QE cross section is strongly suppressed  when $-q^2$ is above 1 GeV, and thus its size notably decreases with $q^0$,  since at the QE-peak, $-q^2 \approx 2Mq^0$. We see that one could only expect to obtain sizable cross sections for forward scattering angles. For instance, at $E_{\nu}=10$ GeV, the energy transfer at the QE peak ranges from very low $q^0\sim 15$ MeV for $\theta=0^\circ$,  up to 7 GeV for $\theta=\theta_{\rm max}(E_\nu=10~{\rm GeV})\sim 25.5^\circ$. Coming back to existence of two QE peaks, the higher one will be much 
more suppressed, and it might not be visible in the differential distribution. For example, at  $E_{\nu}=4$ GeV and in the forward direction, the two peaks occur for $q^2=-0.36$ GeV$^2$ ($q^0= 0.19$ GeV) and $-2.88$ GeV$^2$ ($q^0= 1.53$ GeV), respectively. For larger neutrino energies the $|q^2\,|$ value of the second QE peak grows rapidly, and its impact in the cross sections should become less important.    
%
\begin{figure}[!h]
\centering
\includegraphics[scale=0.35]{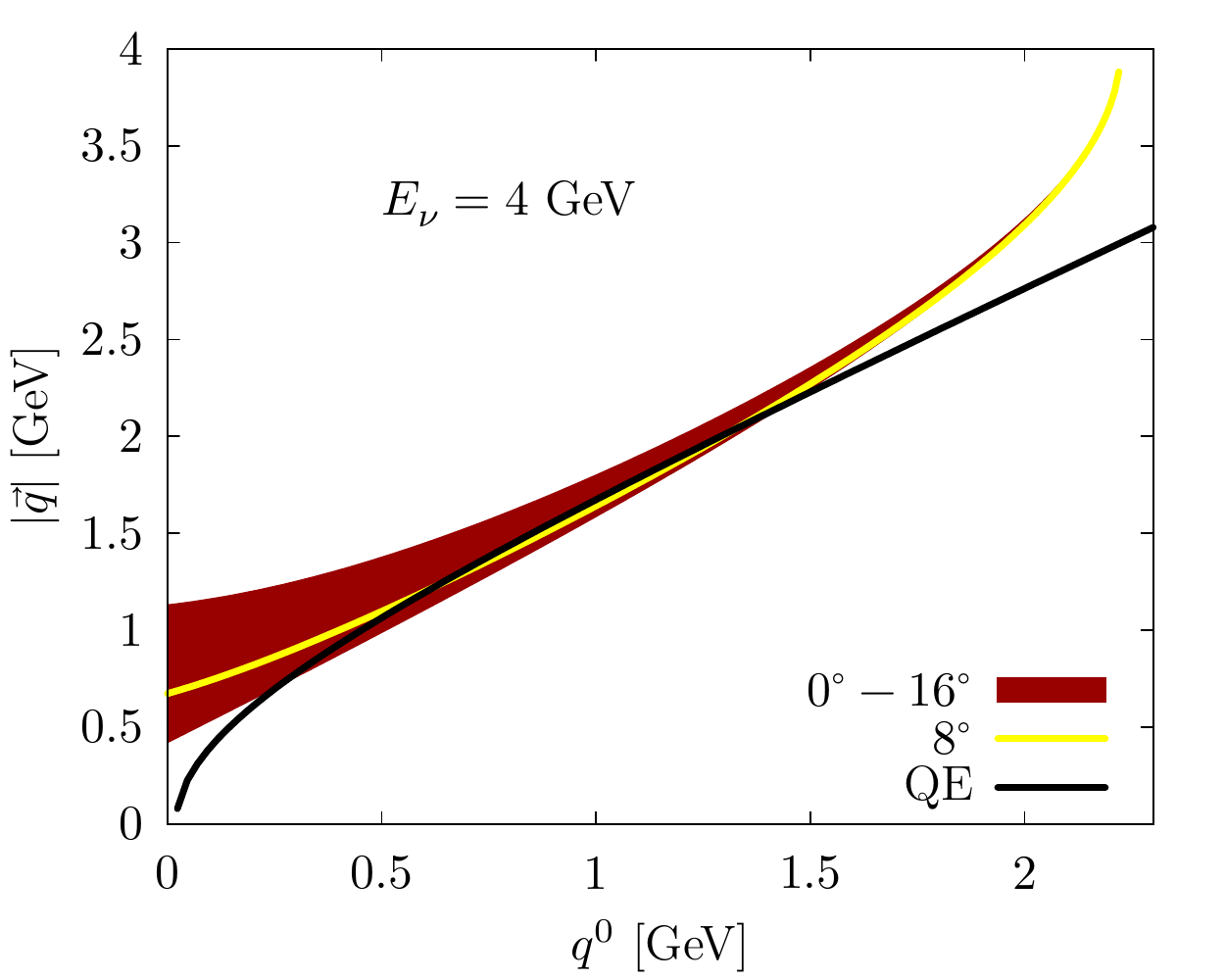}
\includegraphics[scale=0.35]{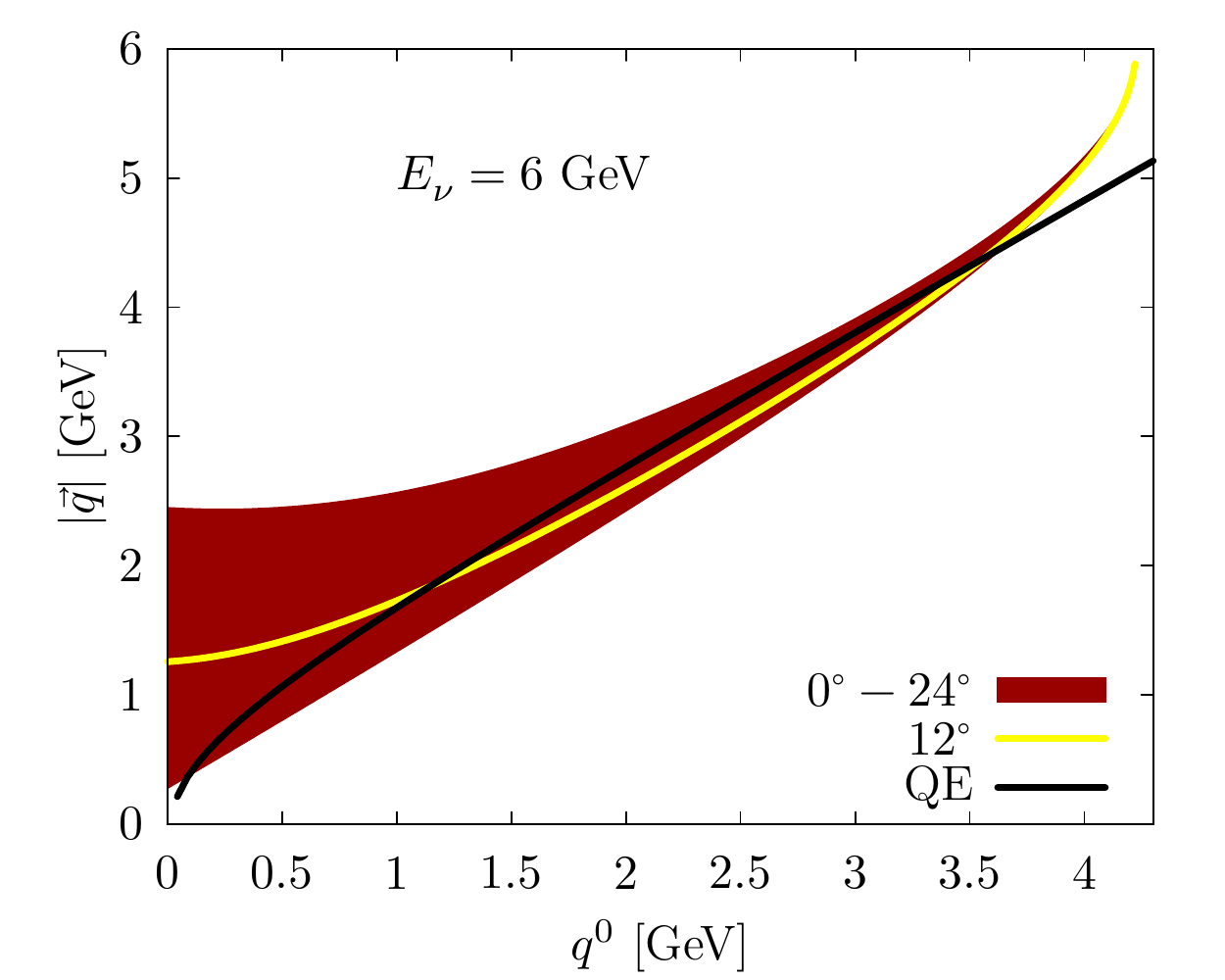}\includegraphics[scale=0.35]{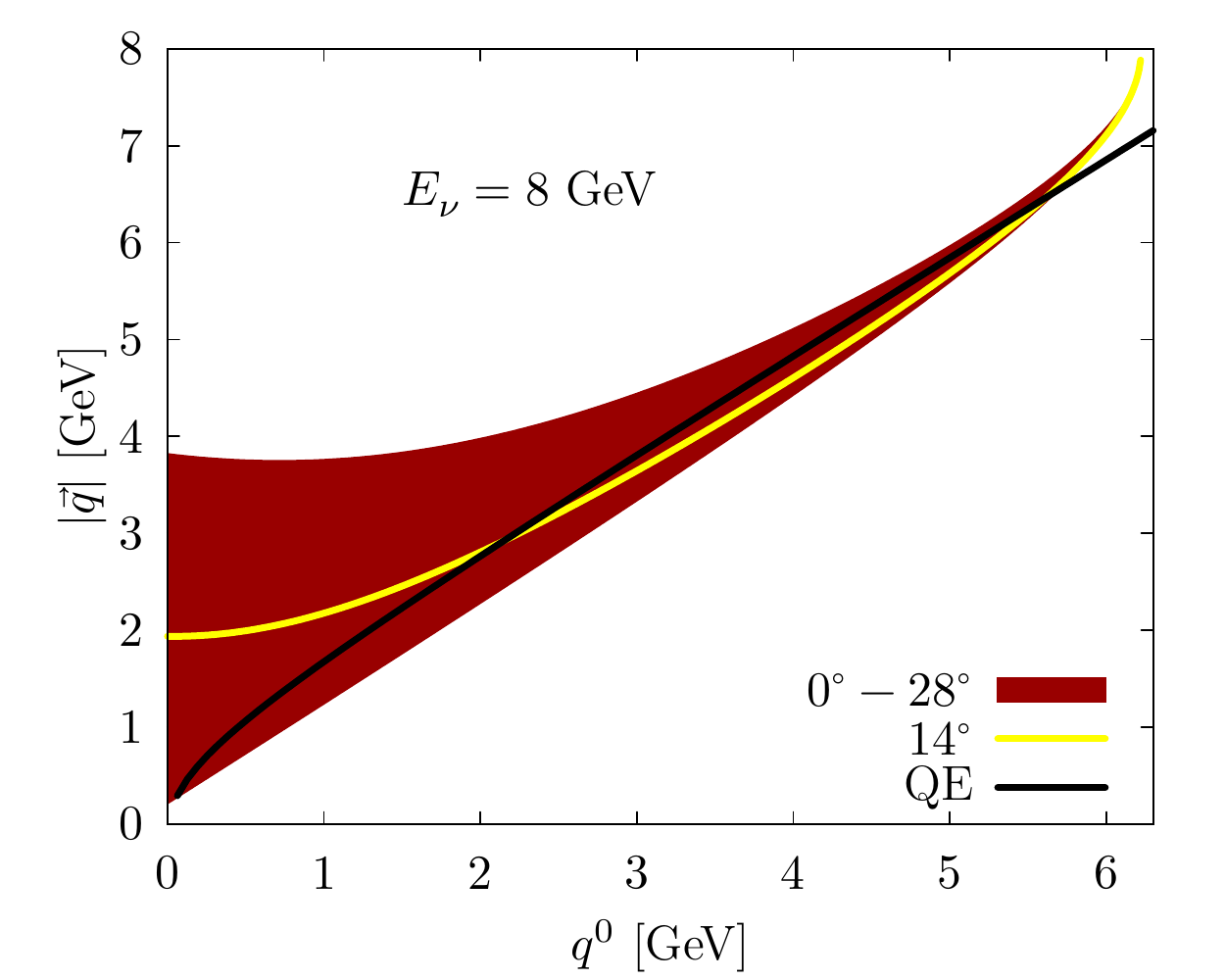}\includegraphics[scale=0.35]{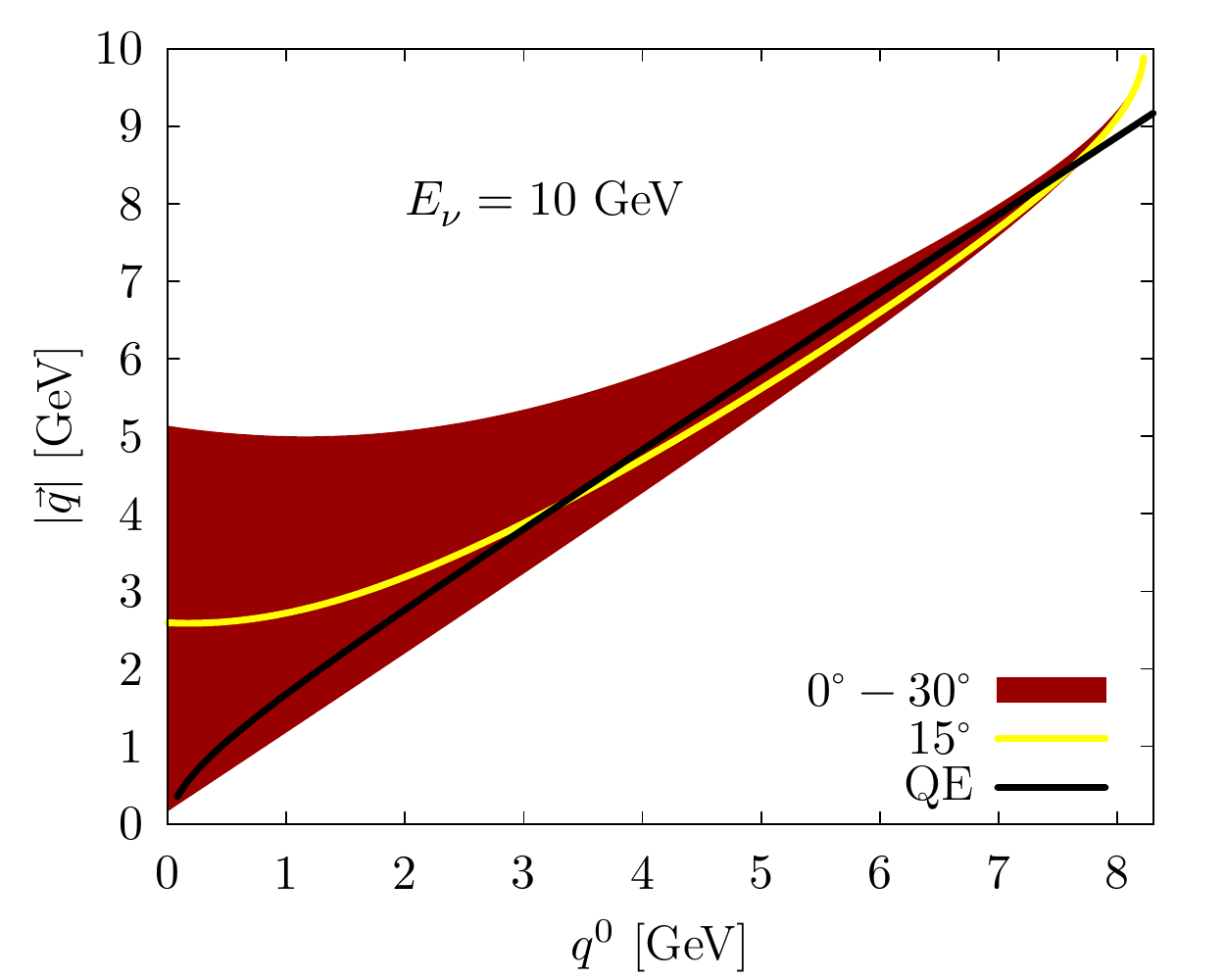}\\
\caption{Some regions of the available phase space for CC $\tau-$ production for $E_{\nu}$= 4, 6 8 and 10 GeV (panels from left to right). We 
display $|\vec{q}\,|_{[E_\nu, q^0, \theta]}$ (Eq.~\eqref{eq:defqtau})  and also the approximated position of the 
QE peak ($|\vec{q}\,|_{\rm QE} = \sqrt{2Mq^0 + q^{02}}$), labeled as QE (black solid line). For each neutrino energy,  the shaded area spans between $\tau-$scattering angle  
$0^\circ$ (lower boundary) up to $16^\circ, 24^\circ, 28^\circ$ and $24^\circ$ (upper boundary), respectively. In each panel, the yellow curve shows $|\vec{q}\,|_{[E_\nu, q^0, \theta]}$ for an intermediate angle among those accounted for in the band. }
\label{fig:phase-space}
\end{figure}
%
Moreover, the results should  be more sensitive to nuclear effects for small values of the scattering angle where the QE cross section is high and peaks in the low $q^0$ region. Hence, we have studied the impact of using the different nuclear SFs to compute the $\nu_\tau/\bar{\nu}_\tau$  
differential cross section and the $\tau$ polarizations in the region of small $\theta$. 
\subsection{Differential cross sections and polarization observables}
%
\begin{figure*}[!h]
\centering
\includegraphics[scale=0.45]{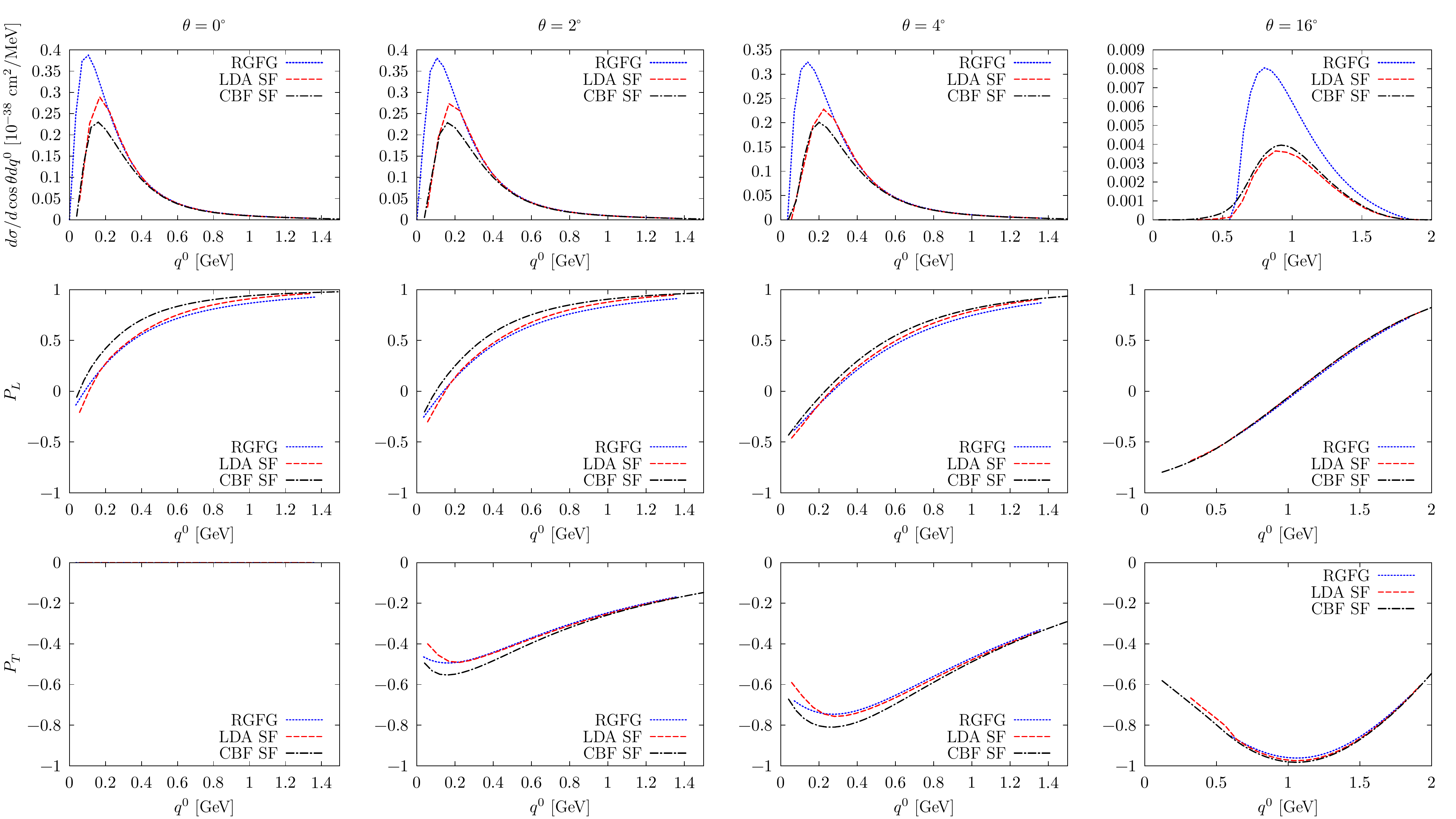}\\
\caption{Double-differential cross section $d\sigma/(d q^0 d\cos\theta)$ and polarization components $P_L$, $P_T$ for $\nu_\tau$ scattering off 
$^{16}$O for $E_\nu=4$ GeV and scattering angles $0^\circ$, $2^\circ$, $4^\circ$ and $16^\circ$.}
\label{fig:4}
\end{figure*}
\begin{figure*}[!h]
\centering
\includegraphics[scale=0.45]{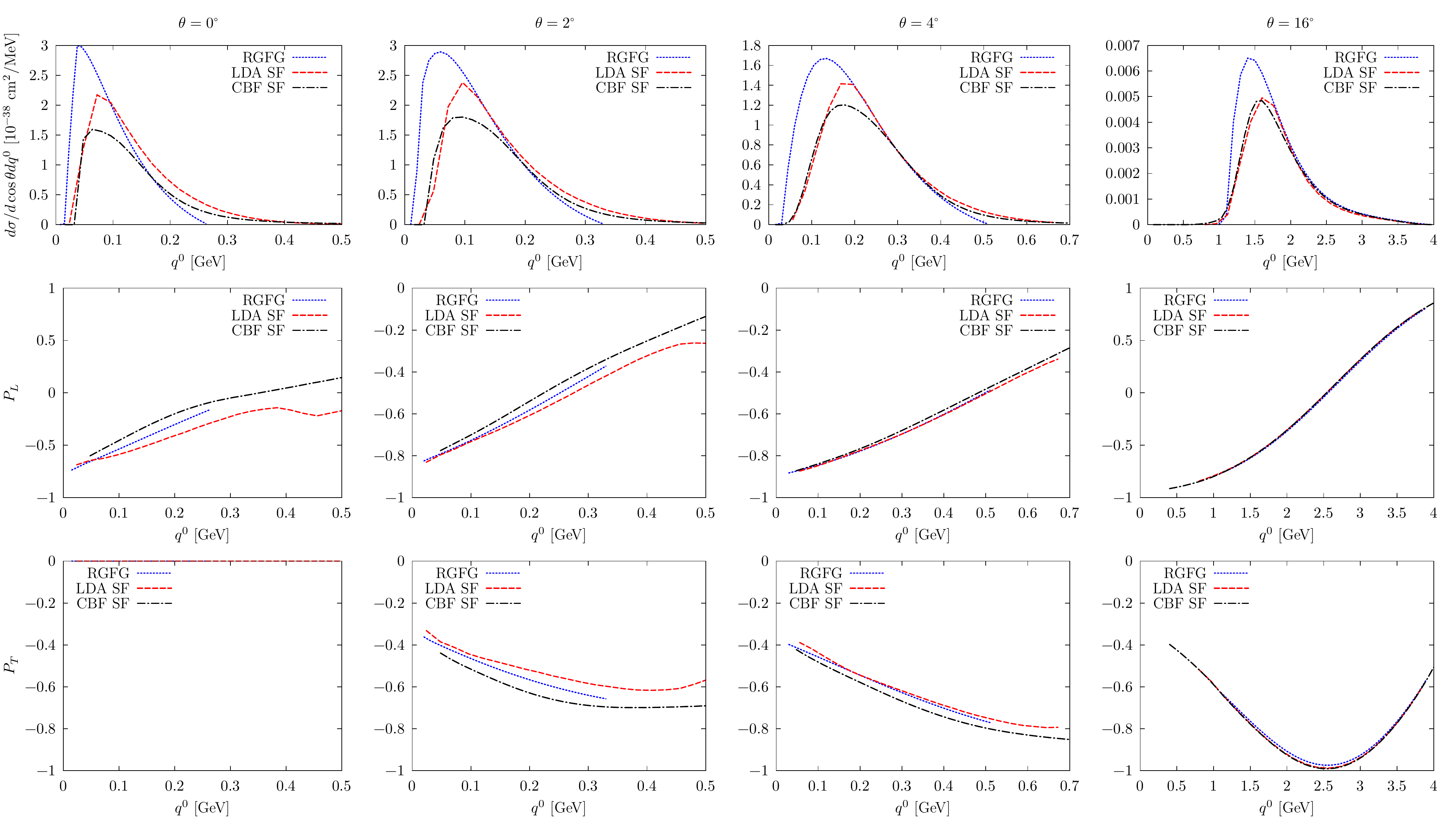}\\
\caption{Same as Fig.~\ref{fig:4} but for $E_\nu=6$ GeV.}
\label{fig:6}
\end{figure*}
%
In Fig.~\ref{fig:4},  we analyze the double-differential cross section (panels in the first row), $P_L$ and $P_T$ (panels in the second and third row, respectively) for the $\nu_\tau +^{16}{\rm O}\rightarrow \tau^- + X$ process at $E_{\nu}= 4$ GeV, and  $\theta=0^\circ, 2^\circ, 4^\circ,$ and $16^\circ$. The dotted (blue) and the  dashed (red) and dot-dashed (black) curves have been obtained using the RGFG model and the LDA and the CBF hole SFs, respectively. Predictions for $E_{\nu}= 6$ GeV are shown in Fig.~\ref{fig:6}. 
The comparison of the three-different sets of results for the differential cross sections clearly reveals that the inclusion of nucleon-nucleon correlations in the hole SF leads to a significant quenching of the QE peak and a shift of its position towards higher energy transfers in  
the dashed (LDA) and dot-dashed (CBF) curves with respect to the dotted one (RGFG).  The distinctive SF tail at high-energy-transfers, 
that arises when short-range correlated pairs are included in the description of the ground state,  is only visible for  $E_{\nu}=6$ GeV. 
Unexpectedly for $E_\nu= 4$ GeV,  the RGFG model seems to provide a $q^0$ tail  similar to those found in the LDA and CBF SF 
calculations. However, its origin in the former model should not be attributed to  short range correlations, but rather it is produced by the  kinematics of the QE CC $\tau-$production. If we come back to Fig.~\ref{fig:phase-space}, we discussed that the $|\vec{q}\,|_{[E_\nu=4\,{\rm GeV}, q^0, \theta=0^\circ]}-$curve intersects the QE-one for two different values of $q^0= 0.19$ and 1.53 GeV. These values correspond to forward and backward $\nu_\tau N \to \tau N$ scattering in the neutrino-nucleon CM system, respectively. The boost to the LAB system converts both CM kinematics into  forward scattering in the LAB frame\footnote{For massless charged leptons, however, the CM backward kinematics does not lead to forward scattering in the LAB frame, while the QE condition for $\theta=0^\circ$ occurs for $q_{\rm QE}^0(m_{\ell}=0)=0$. As the dispersion angle grows, $q^0_{\rm QE} (m_{\ell}=0)$  increases, but there is still a single value where the condition of the QE peak is satisfied (Eq.~\eqref{eq:elmassless}).} (see Appendix~\ref{sec:app}).  In Fig.~\ref{fig:lindhards}, we show the dependence of the  
imaginary part of Lindhard function, ${\rm Im}U(q^0, |\vec{q}\,|_{[E_\nu, q^0, \theta]})$,  on $q^0$ and $q^2$ for charged $\tau$ production for two different kinematics (dashed red curves). As discussed in Ref.~\cite{Nieves:2004wx},  ${\rm Im}U(q)$ essentially gives the  single-nucleon knockout RGFG nuclear response for a unit amplitude,
at the nucleon level. For both kinematics, we clearly see  two peaks,  induced by the forward and backward $\nu_\tau N \to \tau N$ scattering in the neutrino-nucleon CM system, which lead to shapes different from those found for QE-processes involving massless leptons. These distributions should be affected by the nucleon form-factors that produce sizable $q^2-$suppression in the differential cross sections. As an example,  in Fig.~\ref{fig:lindhards} we also display the results modulated by the square of a dipole nucleon form factor, with a cutoff  of 1 GeV$^2$ (dotted blue curves). In these latter cases the second peak disappears, though its existence provides a longer high $q^0-$tail, what is qualitatively observed in the RGFG predictions shown in  Figs.~\ref{fig:4} and ~\ref{fig:6}.
%
\begin{figure*}[!h]
\centering
\includegraphics[scale=0.75]{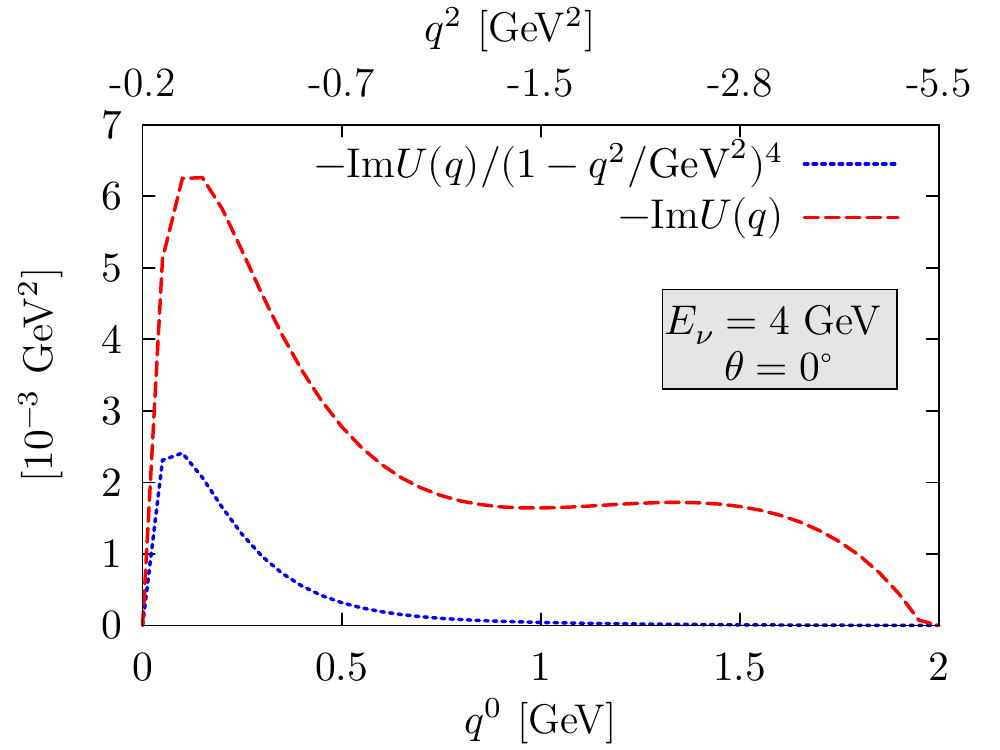}\hspace{1.25cm}\includegraphics[scale=0.75]{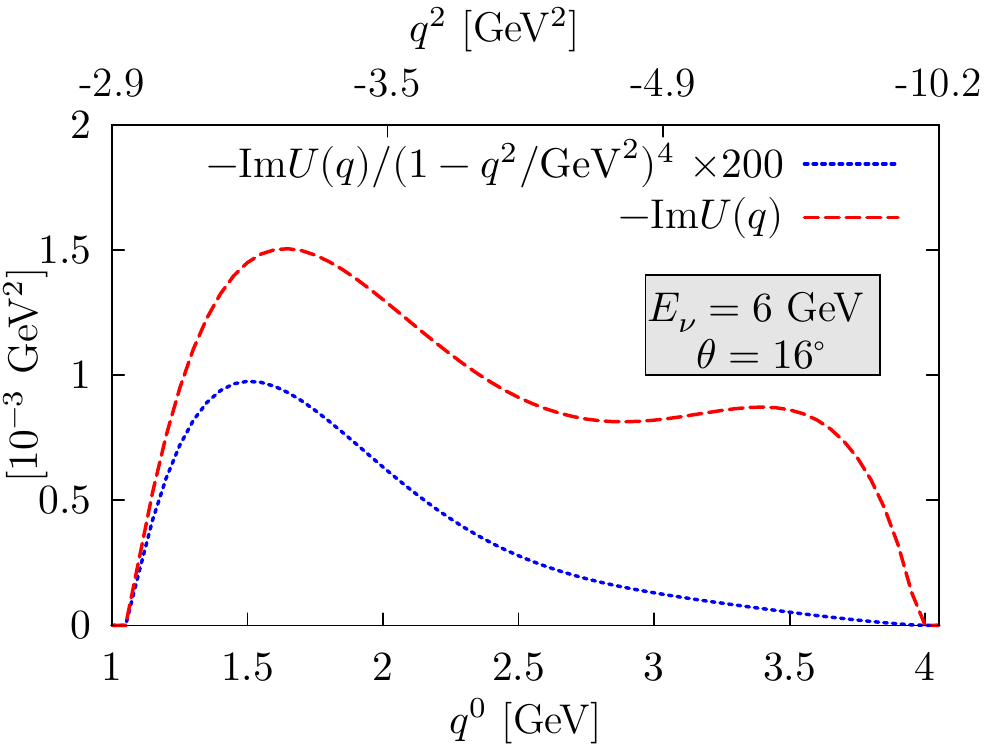}\\
\caption{Imaginary part of the Linhard particle-hole propagator as a function of $q^0$ and $q^2$, with $|\vec{q}\,|= |\vec{q}\,|_{[E_\nu, q^0, \theta]}$ defined in Eq.~\eqref{eq:defqtau}. To evaluate ${\rm Im}U(q^0,|\vec{q}\,|)$ we have used Eq.~(B2) of Ref.~\cite{Nieves:2004wx}, with ${\bar p}_F=0.225$ GeV for both  
neutron and proton Fermi momenta. The left and right panels correspond to $(E_\nu= 4$ GeV, $\theta=0^\circ$) and to   $(E_\nu= 6$ GeV, $\theta=16^\circ$), respectively. In addition, we also show the expected $q^2-$reduction in the cross section provided by dipole weak nucleon form factors, with a mass scale of 1 GeV.}
\label{fig:lindhards}
\end{figure*}

%
Coming back to the discussion of these latter figures, we observe a very nice agreement between the CBF and LDA cross sections for both $E_{\nu}=4$, and 6 GeV and $\theta=16^\circ$ as opposed to $\theta=0^\circ, \ 2^\circ$ and $4^\circ$ cases, where some discrepancies appear. They are likely to be ascribed to the different approximations made to account for the off-shellness of the struck nucleon, as discussed in Subsec.~\ref{sec:hadron_tensor}. These approximations play a more important role in the limit of low momentum transfer and very forward angles.  Let us notice that both SF approaches converge when we move to higher scattering angles, and for $16^\circ$ the differences practically disappear. For the LDA model, the four-momentum $p$ of the initial nucleon is taken from the SF energy--momentum distribution, thus within this scheme, the hole state is treated as an off-shell nucleon. However in the CBF approach, the energy transfer is modified to include the SF effects  
$\tilde{q}^0 \leftrightarrow q^0$, leaving the hole state on-shell (with the momentum taken from the SF and setting $E_p=\sqrt{M^2 + \vec{p}^{\,2}}$). 

The production of the massive $\tau$ lepton is particularly interesting since it might present different polarization components. This fact has a direct implication on the angular distribution of the particles which are subsequently produced in the $\tau$-decay. In the second and third rows on Figs.~\ref{fig:4} and ~\ref{fig:6} we show the impact of nuclear effects on $P_L$ and $P_T$ for different kinematical setups. 
For increasing values of the scattering angle the transverse polarization, $P_T$, of the $\tau$ becomes more visible. This is not surprising because it is proportional to $m_\ell\, \sin\theta$~\cite{Valverde:2006yi}. On the other hand, as the incoming neutrino energy grows, $P_L$ takes values closer to $-1$ for low energy transfers, as expected for the conservation of chirality in CC processes.  One can also observe that $P_L$ and $P_T$ obtained within the CBA and LDA approaches, in most of the cases, do not differ so much from the simplistic RGFG predictions, in spite of leading to significantly distinct  double differential cross sections.  
This should be ascribed to the cancellations that take place when the ratios of Eq.~\eqref{eq:assy} are calculated.  

\begin{figure*}[!h]
\centering
\includegraphics[scale=0.45]{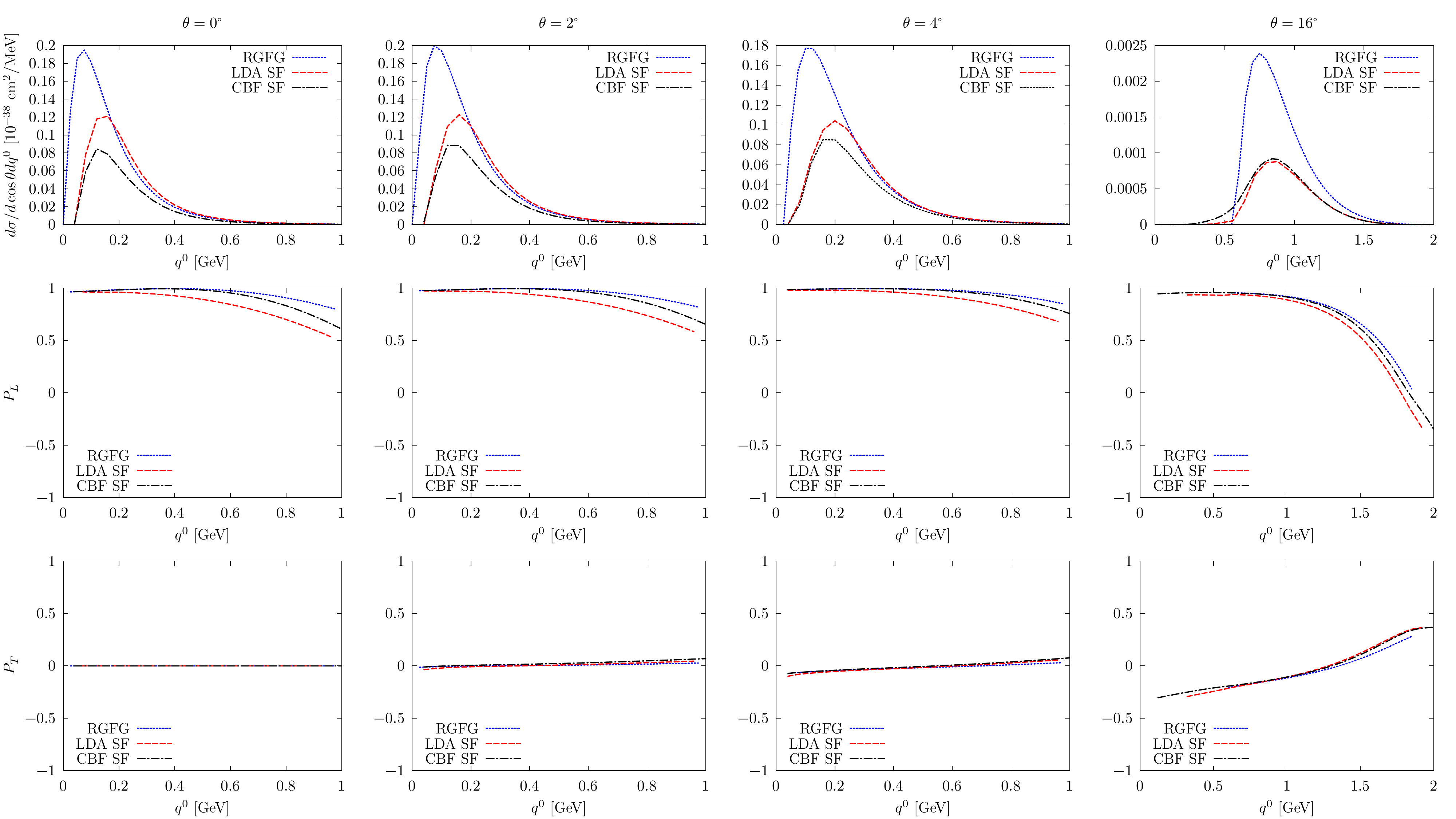}\\
\caption{Double-differential cross section $d\sigma/(d q^0 d\cos\theta)$ and polarization components $P_L$, $P_T$ for $\bar\nu_\tau$ scattering off 
$^{16}$O for $E_\nu=4$ GeV and scattering angles $0^\circ$, $2^\circ$, $4^\circ$ and $16^\circ$.}
\label{fig:anti-4}
\end{figure*}

\begin{figure*}[!h]
\centering
\includegraphics[scale=0.45]{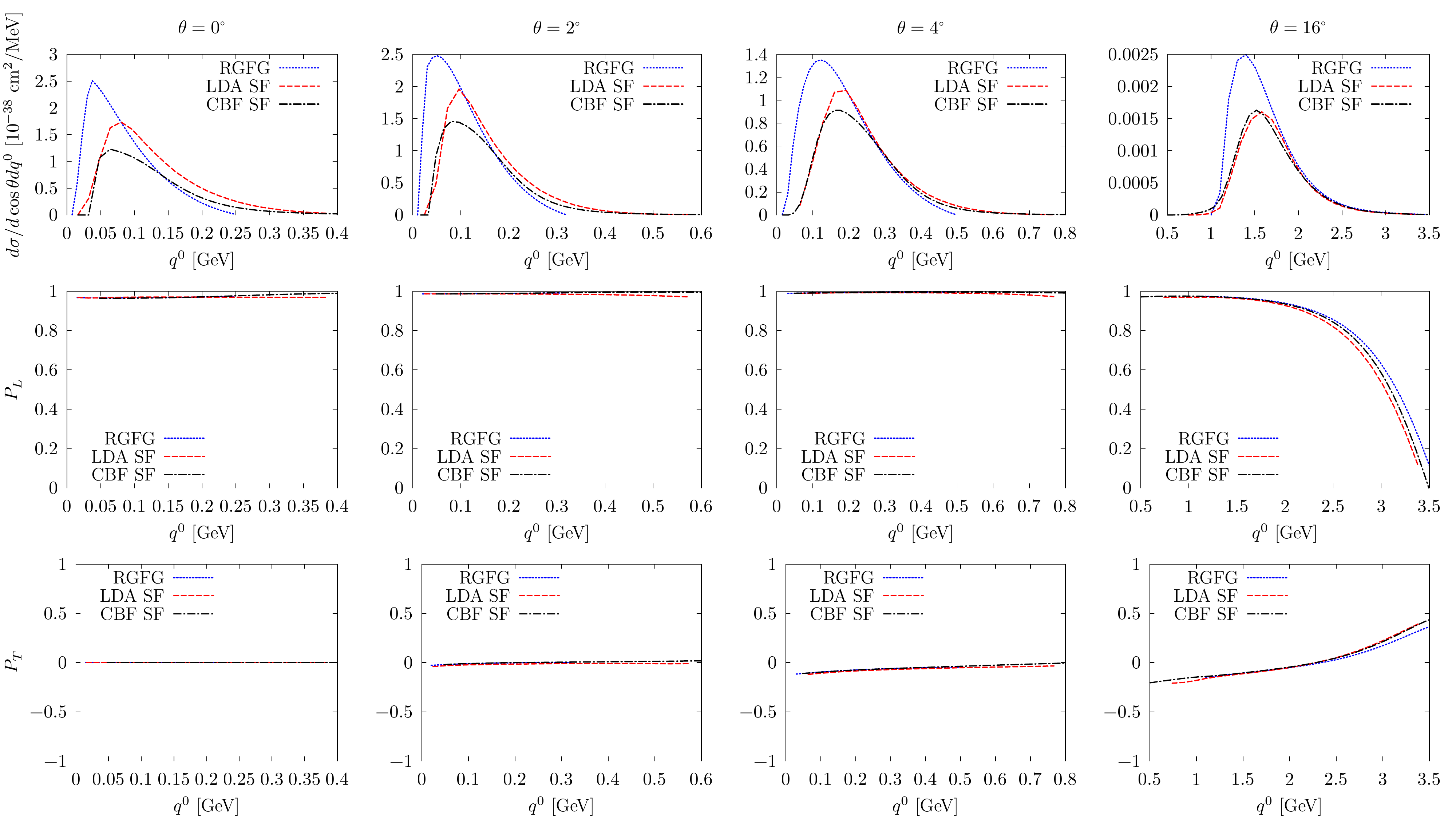}\\
\caption{Same as Fig.~\ref{fig:anti-4} but for $E_\nu=6$ GeV.}
\label{fig:anti-6}
\end{figure*}

Figs.~\ref{fig:anti-4} and \ref{fig:anti-6} are the analogous of Figs.~\ref{fig:4} and \ref{fig:6} but for the $\bar{\nu}_\tau +^{16}{\rm O}\rightarrow \tau^+ + X$ process. The results obtained for the double differential cross sections are qualitatively consistent with the observations we made for the $\nu_\tau$ case. 
Nuclear effects are clearly visible in all the kinematical setups analyzed and the discrepancies between the CBF and LDA predictions are sizable also in this case  up to $\theta=4^\circ$, while for $\theta=16^\circ$ the two set of results are almost coincident. It is worth noticing that the $\tau^{+}$'s  produced in the $\bar{\nu}_\tau$-nucleus scattering are more strongly longitudinally polarized ($P_L\sim 1$ and $P_T\sim 0$) than the $\tau^{-}$'s. 
For $E_{\bar{\nu}}=6$ GeV and scattering angles $0^\circ-4^\circ$, $P_L$ is rather constant through the whole range of available $q^0$ and very close to 1; it departs significantly from 1 only  at $\theta=16^\circ$, when $P_T$ takes larger values (in modulus). For strictly forward scattering,
\begin{eqnarray}
P_L^{(\tau^-, \tau^+)} &= &\mp  \left( 1 - 2\frac{E_{k'}-|\vec{k}\,|}{F^{\nu_\tau, \bar\nu_\tau}(\theta=0^\circ)}\left[ 2W_1 \pm \frac{|W_3|}{M_i}(E_{k'}-|\vec{k}\,|)\right]\right)\nonumber\\ 
&=& \mp  \left( 1 - 2\frac{E_{k'}-|\vec{k}\,|}{M_i\, F^{(\nu_\tau, \bar\nu_\tau)}(\theta=0^\circ)}\left[ W_{xx} \pm |W_{xy}|\right]\right)
\end{eqnarray}
where the factor $F^{\nu, \bar\nu}$ can be found in Eq.~(2) of Ref.~\cite{Valverde:2006yi}, both for neutrino and antineutrino reactions, and $\vec{q}$ is taken in the positive $Z-$direction. There exists a large cancellation between the $xx$ and $xy$ spatial components of the hadron tensor that is responsible for the much smaller values of the antineutrino cross sections than the neutrino ones.   This cancellation also leads to values of  $P_L$ closer to 1 in the case of $\tau^+$ production, and because of the factor $(E_{k'}-|\vec{k}\,|)$, deviations from chirality should increase with the transferred energy $q^0$.  Moreover, differences between nuclear corrections stemming from different approaches should be more visible for antineutrino distributions. Thus for instance within the LDA SF approach, if $p^0$ is fixed to $E_p$ instead of to $E+M$ in Eq.~\eqref{eq:LDA}, values of $P^{\tau^+}_L > 1$ are found in the case of forward antineutrino reactions, while $|P^{\tau^-}_L|$ still keeps smaller than one for neutrino processes. The consistent use of  the energy and momentum distributions obtained from the  hole spectral function $S^{\text{LDA}}_h(E, \vec{p}\,)$, changing the dispersion relation of the initial nucleon, leads to  reasonable predictions for $P^{\tau^+}_L$ below one. 

We observe that the inclusion of the CBF and LDA nuclear SFs significantly  modifies also the  $\bar{\nu}_\tau$-$^{16}$O differential cross section with respect to the RGFG results, and  lead to a significant quenching of the QE peak and a shift of its position 
towards higher energy transfers. The role played by nuclear effects in the determination of $P_L$ and $P_T$ is less systematic. The curves corresponding to the different SFs are found to differ for most of the kinematics considered in Figs.~\ref{fig:4}, \ref{fig:6} ~\ref{fig:anti-4} and \ref{fig:anti-6}. In particular, we find the RGFG predictions for $P_L^{\tau^-}$ and $E_\nu=6$ GeV lay in between the CBF and LDA ones. We interpret this behavior as a manifestation of the strong dependence of the polarization variables on the approximations made in the hadron tensor to treat the off-shell struck nucleon. 

The mean value of the degree of polarization of the $\tau^\mp$ lepton is defined as
\begin{equation}
 \langle {\cal P}^{(\tau^-, \tau^+)} \rangle = \frac{1}{\sigma^{(\nu_\tau, \bar\nu_\tau)}(E_\nu)}\int dE_\tau  d\Omega (\hat{k}') 
 \Sigma_0^{(\nu_\tau, \bar\nu_\tau)}(E_\tau, \theta){\cal P}^{(\tau^-, \tau^+)}(E_\tau, \theta) \label{eq:degP}
\end{equation}
with $\Sigma_0$ the LAB unpolarized double differential cross section of Eq.~\eqref{eq:sec-unpol} and the Lorentz scalar degree of polarization for a given outgoing $\tau-$lepton kinematics given by
\begin{equation}
{\cal P}(E_\tau, \theta)= \sqrt{P_L^2(E_\tau, \theta)+P_T^2(E_\tau, \theta)}
\end{equation}
\begin{figure*}[!h]
\centering
\includegraphics[scale=0.85]{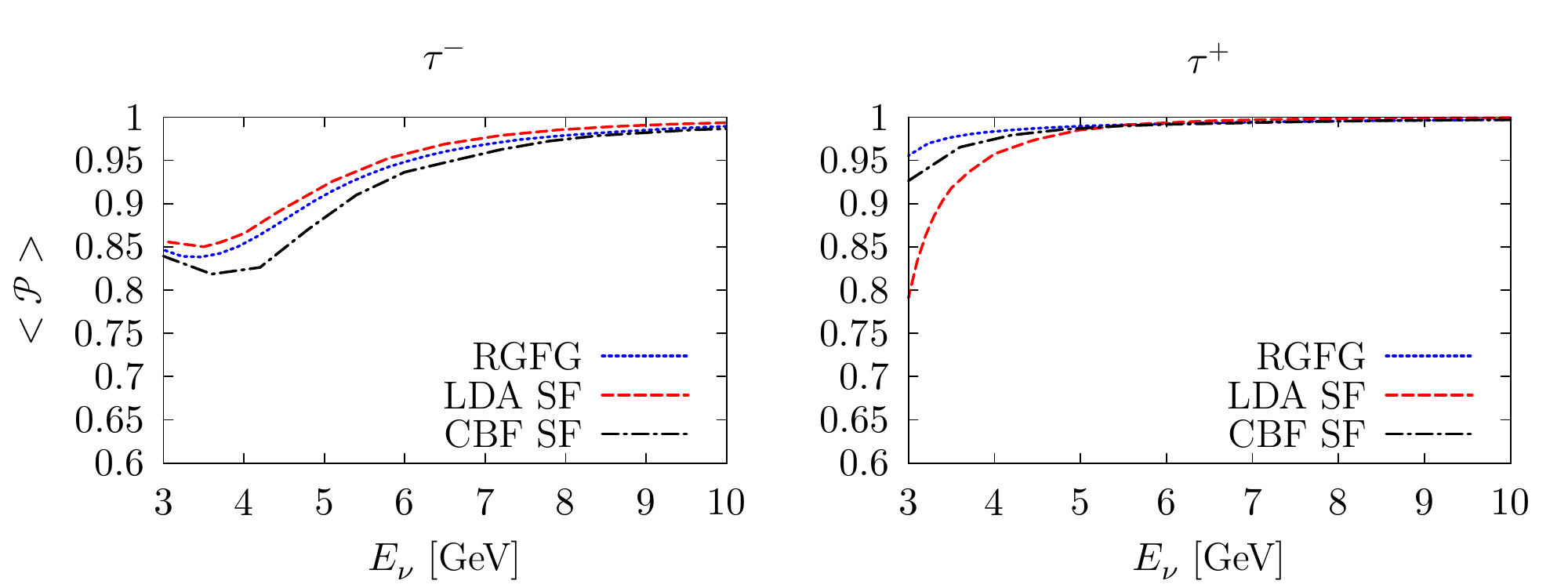}\\
\caption{Mean value of the degree of polarization of $\tau^-$ (left) and $\tau^+$ (right), defined in Eq.~\eqref{eq:degP}, as a function of 
the incoming neutrino energy. We show results for the RGFG model (blue--dotted) and the LDA (red-dashed)  and CBF (black dash-dotted) SF approaches.}
\label{fig:degP}
\end{figure*}
The dependence of the mean value of the degree of polarization  on the neutrino (antineutrino) energy is shown in Fig.~\ref{fig:degP}. The small discrepancies between the curves are likely to originate from the different treatment of the nucleon off-shellness in the LDA and CBF results. However, a clearcut identification of their source can not be easily achieved in this case.  
For the antineutrino, the degree of polarization reaches the asymptotic chiral value more 
rapidly than for the neutrino case,  saturating already at about 5 GeV. Results are in qualitative agreement with those obtained with the simple models considered
in Ref.~\cite{Graczyk:2004uy}. 
%
%
%
%
%
%
%
%
\section{Conclusions}
\label{sec:conclusions}
We have performed an analysis of the cross sections and polarization components for the CC reaction in which a $\nu_\tau/\bar{\nu}_\tau$ scatters off $^{16}$O, focusing on the QE region where the single nucleon knock-out is the dominant reaction mechanism. From the theoretical perspective, the $\nu_\tau$ CC scattering gives a unique opportunity to further investigate the role played by the nuclear correlations of the ground state in the description of neutrino-nucleus interaction. The large mass of the $\tau^\pm$ lepton, with respect to $\mu^\pm$ and $e^\pm$, enables a deeper insight into the nuclear structure of the nucleus when interacting with an electroweak probe. From the total number of five structure functions which are in general needed to describe the hadron tensor of neutrino-nucleus scattering (Eq.~\eqref{eq:hadron_general}), $\nu_e$ and $\nu_\mu$ reactions give access only to three of them, $W_{1,2,3}$, being the contribution of the others suppressed by the low $e^-$ and $\mu^-$  masses.

 The polarization components $P_L$ and $P_T$ of the outgoing $\tau$ are interesting observables, both from the experimental and theoretical point of views. They offer
 deeper insight into the hadron tensor properties, since they are sensitive to different combinations of the structure functions.
In order to provide a realistic description of the nuclear dynamics accounting for  nucleon-nucleon correlations, 
we have used the LDA and CBF hole SFs  derived in  Refs.~\cite{FernandezdeCordoba:1991wf} and ~\cite{Benhar:1989aw, Benhar:1994hw}, successful in modeling inclusive electro-- and (anti)neutrino--nuclear 
QE responses. The SF approaches are substantially more realistic than those based on the use of an effective nucleon mass to describe the initial nuclear state, as done in the previous study of Ref.~\cite{Graczyk:2004uy}. The implementation of a constant effective mass is just a crude approximation to account 
for effects due to the change of the  dispersion relation of a nucleon inside of  a nuclear medium.

For each of the considered models, CBF and LDA, we used slightly different prescriptions of how the single nucleon matrix element is calculated in the nuclear medium. The ambiguity stems from the fact that hit nucleon is off-shell, with its energy-momentum distribution determined by the SF. In the CBF approach the initial nucleon is taken to be on-shell with the momentum distribution taken from the SF. The energy distribution of the SF is taken into account as the modification of the energy transfer. The LDA treats the initial nucleon as an off-shell particle; however, the master equation for the matrix element (where the sum over spins has been performed) is obtained assuming on-shell nucleons. The results for very forward scattering angles turned out to be very sensitive to this choice, affecting both the cross sections and the polarization components. Nevertheless, the discrepancies become very small already for angles $\sim 4^\circ$.
We have shown that the effect of SFs is sizable when the differential cross section is considered, producing a quenching of the QE peak and a shift of its position towards higher energy transfers.  However, nuclear effects are less pronounced for the polarization components, because they are obtained as  ratios of terms proportional to the hadron tensor, where some cancellations occur.

To finish this summary, we would like to recall that RPA correlations  do not change appreciably  the gross features of the polarization of the outgoing $\tau$ charged leptons~\cite{Graczyk:2004uy, Valverde:2006yi}. The reason is
that the polarization components are obtained, as we have just mentioned,  from ratios constructed out of some linear combinations of nuclear structure functions, and the RPA correlations change similarly numerator and denominator.

\begin{acknowledgments}
We warmly thank A. Lovato for useful discussions. We acknowledge the Neutrino Theory Network for partial support.
This research  has been supported by the Spanish Ministerio de Ciencia, Innovaci\'on  y Universidades, European FEDER funds under  Contracts FIS2017-84038-C2-1-P and SEV-2014-0398, and by the EU STRONG-2020 project under the program H2020-INFRAIA-2018-1, 
grant agreement no. 824093.
The work of N.R. is supported by the U.S. DOE, Office of Science, Office of Nuclear Physics, under contract DE-AC02-06CH11357 and by Fermi Research Alliance, LLC, under Contract No. DE-AC02-07CH11359 with the U.S. Department of Energy, Office of Science, Office of High Energy Physics.  \end{acknowledgments}

%
%
%
%
%
%
%
%
\appendix

\section{Lepton polarized CC cross section} 
\label{lept:hadr:terms}
With the axis of quantization in the rest frame [$k'^\mu =(m_\ell, \vec{0}\,)$] of the outgoing $\ell^{\mp}$ specified by the unit vector $\hat n$, we define in this frame a unit space-like four-vector $s^\mu$ as
\begin{equation}
s^\mu = (0, \hat n).
\end{equation}
From the invariance of scalar products, it follows that in any other frame, 
\begin{equation}
 s^\mu = \left (  \frac{\vec{k}'\cdot \hat n}{m_\ell}, \hat n+\frac{\vec{k}'}{m_\ell} \frac{(\vec{k}'\cdot \hat n)}{E_{k'}+m_\ell} \right)\, , \qquad  s^2=-1, \qquad s\cdot k' = 0
 \label{eq:smu}
\end{equation}
The spin-projection operators are then given by~\cite{Mandl:1985bg}
\begin{equation}
\Pi (s;h) = \frac12\left(1+h \gamma_5 \slashed{s} \right), \qquad h= \pm 1\, ,  
\end{equation}
and commute with the energy projection operators $( \pm \slashed{k}'+m_\ell)/2m_\ell $. 

The (anti)neutrino inclusive-differential cross section  for a $(s;h)-$polarized  outgoing lepton is given by (we follow the conventions of Ref.~\cite{Nieves:2004wx}):
\beq\label{eq:1}
\left.\frac{d^2\sigma^{(\nu_\ell, \bar\nu_\ell)}}{d\Omega (\hat{k}') d E_{k'}}\right|_{s;h} =\bigg(\frac{G_F}{2\pi}\bigg)^2 \frac{|\vec{k}'|}{|\vec{k}|} L_{\mu\nu}^{(\nu_\ell, \bar\nu_\ell)}(s;h) W^{\mu\nu}_{(\nu_\ell, \bar\nu_\ell)}
\eeq
in the laboratory (LAB) frame (nucleus at rest). In addition,  $G_F=1.1664\times 10^{-5} \text{GeV}^{-2}$ is the Fermi constant, while $W^{\mu\nu}$ and $L_{\mu\nu}(s;h)$  are the hadronic and polarized lepton tensors, respectively.  The latter tensor reads ($\epsilon_{0123}=+1$)
\begin{eqnarray}
 L_{\mu\nu}^{(\nu_\ell, \bar\nu_\ell)}(s;h) & = & \frac18 {\rm Tr}\left[\slashed{k}'\gamma^\mu \slashed{k}(1+\eta\gamma_5)\gamma^\nu\right] -\eta h  \frac{m_\ell}{8} {\rm Tr}\left[\gamma^\mu \slashed{k}(1+\eta\gamma_5)\gamma^\nu \slashed{s}\right]  \nonumber \\
 &=& \frac{L_{\mu\nu}^{(\nu_\ell, \bar\nu_\ell)}}{2} -\eta h  \frac{m_\ell}{2} s^\alpha \left( k_\mu g_{\nu\alpha}+ k_\nu g_{\mu\alpha}-k_\alpha g_{\mu\nu}+i\eta \epsilon_{\mu\nu\alpha\beta}k^\beta\right)\, ,
 \label{eq:lept_pol}
\end{eqnarray}
where $\eta=\pm$ for the neutrino and antineutrino induced reactions, respectively, $m_\ell$ is the mass of the outgoing lepton, and  $L_{\mu\nu}$  the unpolarized lepton tensor~\cite{Nieves:2004wx}
\beq
\label{eq:lept}
L_{\mu\nu}^{(\nu_\ell, \bar\nu_\ell)} = k'_{\mu} k_{\nu}+k_{\mu} k'_{\nu} - g_{\mu\nu} k\cdot k'+ i\eta\epsilon_{\mu\nu\alpha\beta}k'^{\alpha} k^{\beta}\, ,
\eeq
The hadronic tensor includes all sort of non-leptonic
vertices and corresponds to the charged electroweak transitions of the
target nucleus, $i$, to all possible final states. 
It is thus given by~\cite{Nieves:2004wx}
\begin{eqnarray}
\label{eq:2}
W^{\mu\nu}_ {(\nu_\ell, \bar\nu_\ell)}&=& \frac{1}{2M_i}\overline{\sum_f } (2\pi)^3
\delta^4(P_f-P-q) \langle f | j^\mu_{\rm cc\pm}(0) | i \rangle
 \langle f | j^\nu_{\rm cc\pm}(0) | i \rangle^*
\label{eq:wmunu}
\end{eqnarray}
with $P^\mu$ the four-momentum of the initial nucleus, $M_i=P^2$ 
the target nucleus mass, $P_f$  the total four momentum of
the hadronic state $f$ and $q=k-k^\prime$ the four momentum
transferred to the nucleus.  The bar over the sum denotes the
average over initial spins, and finally for the CC we take
\begin{eqnarray}
j^\mu_{\rm cc+} &=& \overline{\Psi}_u\gamma^\mu(1-\gamma_5)(\cos\theta_C \Psi_d +
\sin\theta_C \Psi_s) \\
j^\mu_{\rm cc-} &=& (\cos\theta_C \overline{\Psi}_d +
\sin\theta_C \overline{\Psi}_s) \gamma^\mu(1-\gamma_5)\Psi_u
\end{eqnarray}
with $\Psi_u$, $\Psi_d$ and $\Psi_s$ quark fields, and $\theta_C$ the
Cabibbo angle ($\cos\theta_C= 0.974$), and $cc\pm$ stand for the neutrino and antineutrino currents.


\section{Kinematics for $\tau$ production off nucleons}
\label{sec:app}
\begin{figure}[htb]
\centering
\includegraphics[scale=0.75]{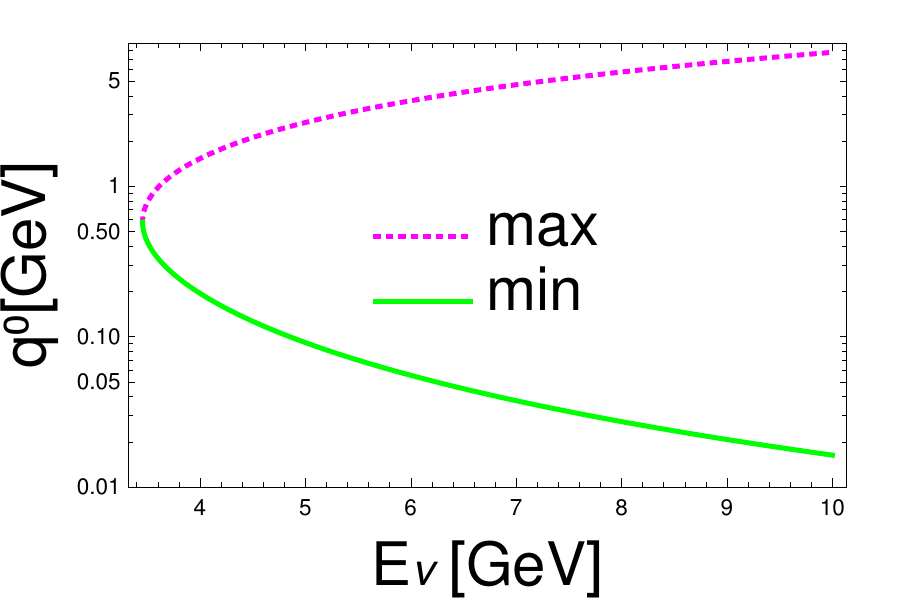}\hspace{1.25cm}\includegraphics[scale=0.75]{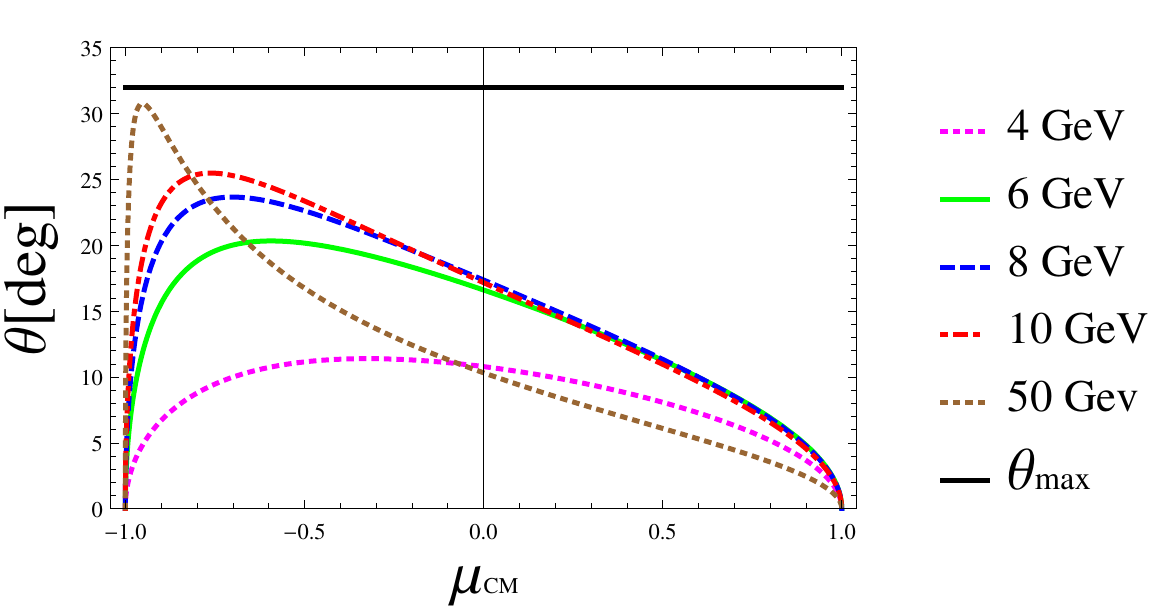}
\caption{Left: Maximum and minimum LAB lepton transferred energies ($q^0_{\rm max, min}= E_\nu-E_{k'}^{\rm min, max}$) as a function of the incoming neutrino energy for weak $\tau$ production off nucleons. Right: LAB $\tau-$lepton scattering angle (deg) as a function of $\mu_{\rm CM}$, or equivalently of the LAB outgoing tau energy, for several incoming neutrino LAB energies: $E_\nu=$ 4 GeV (dotted--magenta), 6 GeV (solid--green), 8 GeV (dashed--blue), 10 GeV (dot-dashed--red) and 50 GeV (dotted--brown). In addition, the horizontal black--solid line stands for the upper bound,  $\theta_{\rm max}= \arcsin\left(M/m_\tau\right)$, that is reached in the  $E_\nu\gg M, m_\tau$ limit.}
\label{fig:thetaLAB}
\end{figure}
%
We will collect here some kinematical relations for the two body reaction
\begin{equation}
 \nu_\ell (E_\nu) N \to \ell(k') N'\ , \quad \ell=e,\mu,\tau
\end{equation}
paying a special attention to the differences induced by the large mass of the $\tau$, with respect to the $\mu-$ and $e-$lepton cases. The LAB threshold neutrino energy, $E_{\nu_\ell}^{\rm th}= m_\ell + m_\ell^2/2M$, is around 3.5 GeV ($\sim 2m_\tau$) for $\tau$ production, while 
the correction ($m_\ell^2/2M$) to $m_\ell$ is negligible for muons and electrons. Taking the incoming neutrino in the positive $Z-$axis,  the lepton scattering angle, $\theta_{\rm CM}$, in the neutrino-nucleon center of mass (CM) frame is not limited  and thus, $\mu_{\rm CM}= \cos\theta_{\rm CM}$ can take any value between $-1$ and 1. The lepton energy ($E_{k'}$) and scattering angle ($\theta$) in the LAB system are obtained through
\begin{eqnarray}
 E_{k'} &=& \frac{(E_\nu+M)E_{k'}^{\rm CM}+\mu_{\rm CM}E_{\nu}|\vec{k}'|^{\rm CM}}{\sqrt{s}} \, , \qquad s = 2ME_\nu + M^2\, , \qquad E_{k'}^{\rm CM}= \frac{s+m^2_\ell-M^2}{2\sqrt{s}}\label{eq:elepton}\\
 \tan\theta &=& \frac{\sqrt{s}}{E_\nu+M} \frac{\sqrt{1-\mu_{\rm CM}^2}}{\mu_{\rm CM}+a} \, , \qquad a = \frac{E_\nu}{E_\nu+M}\frac{E_{k'}^{\rm CM}}{|\vec{k}'|^{\rm CM}} \label{eq:tan}
\end{eqnarray}
with $|\vec{k}'|^{\rm CM}= \sqrt{(E_{k'}^{\rm CM})^2-m^2_\ell}$. The maximum and minimum LAB energies of the  outgoing lepton correspond to $\mu_{\rm CM}=+1$ and $-1$, respectively, and they read
\begin{equation}
 E_{k'}^{\rm max} = \frac{(E_\nu+M)E_{k'}^{\rm CM}+E_{\nu}|\vec{k}'|^{\rm CM}}{\sqrt{s}} \, , \qquad E_{k'}^{\rm min} = \frac{(E_\nu+M)E_{k'}^{\rm CM}-E_{\nu}|\vec{k}'|^{\rm CM}}{\sqrt{s}} \label{eq:ene-range}
\end{equation}
The range of transferred energies to the nucleon in the LAB system, and for $\tau-$lepton production, is shown in the left panel of Fig.~\ref{fig:thetaLAB} 
as a function of $E_\nu$, up to 10 GeV.

The parameter $a>0$, introduced in Eq.~\eqref{eq:tan}, plays an important role to determine the LAB angular distribution. It diverges at threshold and it becomes 1 when $E_\nu\gg M, m_\ell$. 
For muon production $a<1$, except for a very narrow region ($\sim 1$ MeV) comprised between threshold and $E_\nu^{a=1}= m_\mu (M-m_\mu/2)/(M-m_\mu)$, with $m_\mu$ the muon mass\footnote{The parameter $a$ takes the value of 1 for $ E_\nu = E_\nu^{a=1} $, reaches a minimum above this energy and after, it begins to grow approaching the asymptotic value of 1 for large neutrino energies.}. The situation, conveniently re-scaled, is similar for  electron production. For $\tau$ production, however, $a$ is a decreasing monotone function, being always greater than 1 and  reaching this latter value only in the $E_\nu\to \infty$ limit. Thus, we have
\begin{itemize}
 \item $a<1$ (muon and electron production):  The LAB lepton scattering angle $\theta$ can take any value between 0 and $\pi$, with $\theta > \pi/2$ ($\theta \le \pi/2$) for $\mu_{\rm CM} < -a$ ($\mu_{\rm CM} \ge -a$). Furthermore, there is a biunivocal correspondence between $\mu_{\rm CM}$ and  
 $\cos\theta$,  and hence between the LAB variables $E_{k'}$ and $\cos\theta$. Namely, neglecting the muon or electron masses with respect to that of the nucleon or the neutrino energy, one finds
 \begin{equation}
  E_{k'} = \frac{M E_\nu }{M+E_\nu- E_\nu\cos\theta} \label{eq:elmassless}
 \end{equation}
 \item $a>1$ (tau production):  We see that $\tan\theta$ is always greater than zero, and therefore $\theta< \pi/2$. Actually, $\tan\theta$, seen as a function of $\mu_{\rm CM}$ or equivalently of the LAB outgoing tau energy, has a maximum for $\mu_{\rm CM}= -1/a< 0$. We find that the maximum lepton scattering angle in the LAB frame, $\theta_{\rm max}$, is
\begin{equation}
 \theta_{\rm max} = \arcsin\left(\frac{\sqrt{s}|\vec{k}'|^{\rm CM}\,}{E_\nu m_\tau}\right) < \arcsin\left(\frac{M}{m_\tau}\right)\sim 31.9^o \label{eq:thetamax}
\end{equation}
where the upper bound is reached for  $E_\nu\gg M, m_\tau$. The dependence of $\theta$ on $\mu_{\rm CM}$ is shown in the right panel of Fig.~\ref{fig:thetaLAB} for different incoming neutrino LAB energies. We observe that any LAB $\tau-$scattering angle is obtained for two  different values of $\mu_{\rm CM}$ (two different values of the LAB $\tau-$energy, $E_{k'}$, as inferred from  Eq.~\eqref{eq:elepton}), and hence the $(\theta,\mu_{\rm CM})-$correspondence is not biunivocal in this case. One of the solutions (A) always corresponds to the $\tau-$lepton coming out backwards in the CM frame ($\mu_{\rm CM }< -1/a \leftrightarrow \theta_{\rm CM} > \pi/2$). For the second one (B) $\mu_{\rm CM }> -1/a$, which depending on the neutrino energy and on $\theta$ might also correspond to $\mu_{\rm CM }< 0$. The CM to LAB boost transforms both CM configurations into quite forward scattering in the LAB system ($\theta < 32^o$). The B--solution gives rise to a larger (smaller) outgoing LAB $\tau-$energy (transferred energy $q^0=E_\nu-E_{k'}$) than the A one. The details of the 
$\theta(\mu_{\rm CM})$ distribution depends on the incoming neutrino energy, as can be seen in Fig.~\ref{fig:thetaLAB}, and its asymmetry becomes more pronounced as $E_\nu$ grows, with the the maximum position approaching to $\mu_{\rm CM}=-1$ and $\theta$ at the maximum to $\theta_{\rm max}$.

This is the kinematics that always applies for $\tau-$production, while as mentioned for muon or electron weak production,  the parameter $a$ is greater than zero only for a very narrow  range of LAB neutrino energies comprised between  $m_\ell + m_\ell^2/2M$ and $m_\ell (M-m_\ell/2)/(M-m_\ell)$, with $\ell=e,\mu$.

\end{itemize}

\bibliography{biblio}

\end{document}